\documentclass[11pt]{article} % use larger type; default would be 10pt
\usepackage[utf8]{inputenc} % set input encoding (not needed with XeLaTeX)
\usepackage[dvips,letterpaper,margin=0.8in]{geometry}
\usepackage{amsmath}
\usepackage{amsfonts}
\usepackage{amssymb}
\usepackage{graphics}
\usepackage{graphicx}
\usepackage{amssymb}
\usepackage{cite}
\usepackage{setspace}
\usepackage{appendix}
\usepackage{graphicx} % support the \includegraphics command and options

\usepackage{booktabs} % for much better looking tables
\usepackage{array} % for better arrays (eg matrices) in maths
\usepackage{paralist} % very flexible & customisable lists (eg. enumerate/itemize, etc.)
\usepackage{verbatim} % adds environment for commenting out blocks of text & for better verbatim
\usepackage{subfig} % make it possible to include more than one captioned figure/table in a single float
\usepackage{fancyhdr} % This should be set AFTER setting up the page geometry
\usepackage{sectsty}
\allsectionsfont{\sffamily\mdseries\upshape} % (See the fntguide.pdf for font help)
\usepackage[nottoc,notlof,notlot]{tocbibind} % Put the bibliography in the ToC
\newcommand{\inv}{^{-1}}                                               %     Inverse symbol
\newcommand{\ra}{\rightarrow}                                        %     Single arrow to the right
\newcommand{\N}{\mathbb{N}}                                        %     Natural Numbers
\newcommand{\beq}{\begin{equation}}
\newcommand{\eeq}{\end{equation}}
\newcommand{\bs}{\boldsymbol}
\numberwithin{equation}{section}

\title{Unitary representations of the Galilean line group: \\ Quantum mechanical principle of equivalence}
\author{B. R.~MacGregor, A. E.~McCoy and S.~Wickramasekara\\
Department of Physics\\
Grinnell College\\
Grinnell, IA 50112}
\date{} 

\begin{document}
\maketitle
\begin{abstract}
We present a formalism of Galilean quantum mechanics in non-inertial reference frames and discuss its implications for the equivalence principle. 
This extension of  quantum mechanics rests on the Galilean line group, the semidirect product of the 
real line and the group of analytic functions from the real line to the Euclidean group in three dimensions. This group provides transformations between 
all inertial and non-inertial reference frames and contains the Galilei group as a subgroup. We construct a certain class of unitary representations 
of the Galilean line group and show that these representations determine  the structure of quantum mechanics in non-inertial reference frames. 
Our representations of the Galilean line group contain the usual unitary projective representations 
of the Galilei group, but have a more intricate cocycle structure. The transformation formula for the Hamiltonian under the Galilean line group shows that in a non-inertial reference frame 
it acquires a fictitious potential energy term that is proportional to the inertial mass, suggesting the equivalence of inertial mass and gravitational mass in quantum mechanics. 
\end{abstract}

\section{Introduction}
The purpose of this paper is to present a formulation of quantum mechanics in non-inertial reference frames. The profound connection between non-inertial 
reference frames and classical gravity, encapsulated in the principle of equivalence and  brought to light fully by Einstein's magnificent theory, 
suggests that a formulation of quantum mechanics in non-inertial reference frames should shed light on the question of whether the principle 
of equivalence is compatible with the axioms of quantum mechanics. 

Inertial reference frames have a long and interesting history. Newton clearly recognized that his laws of mechanics, as they are stated in 
 \emph{Principia}, hold only  in a special class of reference frames. For instance, an isolated particle has zero acceleration only in these frames. 
 For Newton, these were the reference frames that were either at rest or moving with constant velocity with respect to his absolute space. 
 Philosophers continued to debate the issue; Leibniz and Berkeley, among others,  found the notion of absolute space abhorrent, while Euler and Kant, among others, sided with Newton. From a practical standpoint, the important point about the absolute space is that it asserts the \emph{existence} 
 of inertial reference frames. It is as such an assertion that Newton's first law acquires content, rather than being reduced to a special case of his second law. 
 
An early advocate of positivism, Mach concerned himself with finding an \emph{operational definition} of inertial reference frames.  He recognized the  
relationship between inertia and gravitation and argued that the inertia of a particle is the manifestation of the net gravitational effect of 
all the matter in the universe on that particle.  As is well known, the views of Mach had a strong impact on Einstein. 
From the observation that the effect of a uniform gravitational field 
on a particle completely disappears when viewed from a reference frame freely falling in that gravitational field, Einstein reasoned 
that all laws of physics as stated in a freely falling reference frame and in an inertial reference frame are  indistinguishable in form. That is, 
inertial reference frames are those in free fall in the ambient gravitational field - his ``happiest thought."  Introducing a notion of locality 
to accommodate nonuniform gravitational fields, Einstein constructed his general theory of relativity on the core physical principle 
that for any given point in spacetime, there exists \emph{local} reference frames in which all laws of physics have the same form as they do in inertial reference frames. 
Infinitely many such local inertial reference frames are clearly possible at any given spacetime point. In Einstein's theory, they
 are taken to be 
related to one another by Lorentz transformations so that general relativity would be consistent with his special theory of relativity. However, it was understood 
early on that general covariance by itself is devoid of physics content and that it is equally possible in a generally covariant theory to demand that local inertial 
reference frames be transformed amongst themselves by Galilean transformations~\cite{kretschmann}.  
On the other hand, the covariance  under transformations that connect different local inertial reference frames, be they Lorentz transformations or Galilean transformations, 
imposes stringent restrictions on possible laws of physics. 

The requirement that the laws of physics be covariant in all inertial reference frames, often called the principle of relativity, also plays 
a crucial role in quantum mechanics. In fact, the principle of relativity and the principle of superposition are the two 
fundamental physical principles on which quantum theory rests. To a large extent, the structure of quantum theory, both relativistic and non-relativisitc, is 
determined by these two principles. 

To be specific, the synthesis of the principles of superposition and relativity leading to quantum mechanics 
can be achieved by way of representations of the relevant symmetry group of spacetime -- in particular, 
the Galilei group or the Poincar\'e group for non-relativistic or relativistic quantum physics, respectively.  The vector space that furnishes the representation is interpreted as 
the state space of the quantum system. 
Supplementary requirements further constrain these representations. For instance, the requirement that the probability amplitude be given by an inner product necessitates 
that the representation space be an inner product space (leading to a Hilbert space) and that the representation be unitary.  Well-known features  of quantum theory such 
as Schr\"odinger's equation and Heisenberg's canonical commutation relations follow as consequences of these unitary representations of spacetime symmetry groups.  

In this paper, we seek to advance this view that the structure of quantum mechanics is determined by the 
unitary representations of the relevant spacetime symmetry group and to generalize the notion of a spacetime 
symmetry  group to include  non-inertial reference frames.  We will limit ourselves to non-relativistic quantum physics, 
i.e., we take the subset of frames consisting of  inertial reference frames to be closed under the action of the Galilei group. Non-intertial reference 
frames that we consider are precisely the set of reference frames that is closed under the action of the semidirect product of the real line and 
the (infinite dimensional) 
group of analytic functions from the real line to the Euclidean group in three dimension. We will refer to this group  as the 
\emph{the Galilean line group}. 
Hence, our approach bears a certain resemblance to gauge 
field theories, which carry a representation of the group 
of functions from $\mathbb{R}^4$ to a compact group (furnishing transformations of certain internal degrees of freedom). 
We will then construct a class of  unitary representations of the Galilean line group so that they contain unitary, irreducible, projective 
representations of the Galilei group. The main technical result we report in this paper is the construction of these 
representations. Of particular interest is the property that, although 
the Galilei group is naturally contained in its line group,  there exist no 
central extensions of the Galiean line group that contain a given central extension of the Galilei group. However, as shown in section \ref{sec3.4}, 
there does 
exist a certain non-central extension of the Galilean line group that contains a given central extension of the Galilei group.  This 
non-central extension of the line group is identical to the line group of the centrally extended Galilei group.  Our main assertion is that it is precisely the unitary 
representations of the (non-centrally) extended Galilean line  group that underlies what may be called the quantum mechanical equivalence 
principle.  

We have come to know of one other paper \cite{klink} that takes a similar approach to understanding the role equivalence principle in non-relativistic quantum mechanics. 
The conclusions of \cite{klink} are rather similar to ours. The main difference between \cite{klink} and the study presented here is that time translations are treated separately 
in \cite{klink} as a way of circumventing the problem of the absence of good  central extensions of the line group. In contrast, we treat time translations in a more integrated 
manner as a part of the line group. As pointed out above, this necessitates the construction of certain non-central extensions of the Galilean line group. As a result, our 
representations are different from that of \cite{klink}, particularly with regard to their cocycle structure.  

The organization of the paper is as follows: in Section \ref{sec2}, we will present a brief review of the Galilei group, its unitary representations and 
its central extensions.  In Section \ref{sec3}, we will introduce the Galilean line group and its algebra and construct its extensions. Section \ref{sec4} will be devoted to the 
construction of  unitary representations of the extended Galilean line group. In particular, we will show that when the Hamiltonian is transformed under the line group, 
it acquires additive terms that are dependent on the  acceleration of the transformed frame.  Moreover, these fictitious potential energy terms are all proportional to the inertial mass, just as 
one would expect from the classical equivalence principle. In Section \ref{sec5}, we will discuss the role of the equivalence principle in quantum mechanics and 
make some concluding remarks.

\section{The Galilei group}\label{sec2}
Galilei group is the set ${\cal G}=\{(R,\bs{v},\bs{a},b)\}$, 
along with the mapping ${\cal G}\otimes{\cal G}\to{\cal G}$ defined by 
\begin{equation}
(R_2,\bs{v}_2,\bs{a}_2,b_2)(R_1,\bs{v}_1,\bs{a}_1,b_1)=(R_2R_1,\bs{v}_2+R_2\bs{v}_1,\bs{a}_2+R_2\bs{a}_1+b_1\bs{v}_2,b_2+b_1).\label{2.1}
\end{equation}
Here, the $R$ are orthogonal matrices, $\bs{v}$ and $\bs{a}$ are vectors under $R$, and $b$ is a scalar. 
The associativity of \eqref{2.1} can be readily verified. Under \eqref{2.1}, each element of ${\cal G}$ has an inverse, given by
\begin{equation}
(R,\bs{v},\bs{a},b)^{-1}=(R^{-1},-R^{-1}\bs{v}, -R^{-1}(\bs{a}-b\bs{v}),-b).\label{2.2}
\end{equation}
The Galilei group has realization as a transformation group on the four dimensional spacetime $\mathbb{R}^3\otimes\mathbb{R}$ given by the formula
\begin{equation}
(R,\bs{v},\bs{a},b):\quad \left(\begin{array}{c}
\bs{x}\\
t
\end{array}
\right)\to\left(\begin{array}{c} \bs{x}' \\ t'\end{array}\right)  = \left( \begin{array}{c} 
R\bs{x} + \bs{v}t + \bs{a} \\
t+b\end{array}\right).\label{2.3}
\end{equation}
With \eqref{2.3}, the group parameters $R,\ \bs{v},\ \bs{a}$ and $b$ acquire interpretation, respectively, as rotations in $\mathbb{R}^3$, Galilei boosts, 
space translations and time translations.  The Galilei group ${\cal G}$ can also be written as a group of $5\times5$ matrices by writing \eqref{2.3} in the following 
form:
\begin{equation}
\left(\begin{array}{c}
\bs{x}'\\
t'\\
1
\end{array}
\right)=
\left(\begin{array}{ccc}
R&\bs{v}&\bs{a}\\
0&1&b\\
0&0&1
\end{array}
\right)
\left(\begin{array}{c}
\bs{x}\\
t\\
1
\end{array}
\right).\label{2.4}
\end{equation}

It follows from \eqref{2.3} or \eqref{2.4} that there exits a representation $\Phi\times{\cal G}\to\Phi$ on a suitably defined vector space 
$\Phi$ of real or complex functions $f$ on the spacetime manifold $\mathbb{R}^3\otimes\mathbb{R}$. 
The action of the group operators $\hat{U}\left(R,\bs{v},\bs{a},b\right)$ furnishing the representation is 
defined by 
\begin{eqnarray}
\left(\hat{U}(R,\bs{v},\bs{a},b)f\right)(\bs{x},t)&=&f\left((R,\bs{v},\bs{a},b)^{-1}(\bs{x},t)\right)\nonumber\\
&=&f\left(R^{-1}\bs{x}-R^{-1}\bs{v}t-R^{-1}(\bs{a}-b\bs{v}),t-b\right).
\label{2.5}
\end{eqnarray}
If we let  $\Phi$  be the Hilbert space $L^2(\mathbb{R}^4)$ of complex-valued functions with the usual Lebesgue measure, 
then the representation \eqref{2.5} is unitary. We will take this to be the case in the following. 

Let $i\left.d\hat{U}(\cal G)\right|_{e}$ be the algebra of differentials of  \eqref{2.5}  evaluated at the identity 
$e$ of the Galilei group. A basis for this operator algebra can be chosen to consist of the following: 
\begin{subequations}\label{2.6}
\begin{eqnarray}
\hat{J}_if(\bs{x},t)&:=&i\frac{\partial}{\partial \theta}\left.\left(\hat{U}(R_i(\theta),\bs{0},\bs{0},b)f\right)\right|_{e}(\bs{x},t)\nonumber\\
&=&-i\epsilon_{ijk}x^j\frac{\partial}{\partial x^k}f(\bs{x},t)\label{2.6a}\\
\hat{K}_if(\bs{x},t)&:=&i\frac{\partial}{\partial v_i}\left.\left(\hat{U}(I,{v_i},\bs{0},0)f\right)\right|_{e}(\bs{x},t)\nonumber\\
&=&-it\frac{\partial}{\partial x^i}f(\bs{x},t)\\
\hat{P}_if(\bs{x},t)&:=&i\frac{\partial}{\partial a_i}\left.\left(\hat{U}(I,\bs{0},{a_i},0)f\right)\right|_{e}(\bs{x},t)\nonumber\\
&=&-i\frac{\partial}{\partial x^i}f(\bs{x},t)\\
\hat{H}f(\bs{x},t)&:=&i\frac{\partial}{\partial b}\left.\left(\hat{U}(I,\bs{0},\bs{0},b)f\right)\right|_{e}(\bs{x},t)\nonumber\\
&=&-i\frac{\partial}{\partial t}f(\bs{x},t)\label{2.6e}
\end{eqnarray}
\end{subequations}
Our notation in \eqref{2.6a} is that $R_i(\theta)$ denotes a counterclockwise rotation about the $x_i$-axis by an angle $\theta$. 

It should be noted that the limits involved in the derivatives of \eqref{2.6} are to be taken with respect to the norm topology of 
$L^2(\mathbb{R}^3\otimes\mathbb{R})$. For instance, the last equality defining the Hamiltonian $\hat{H}$ has content as the $L^2$-limit 
\begin{eqnarray}
\lim_{\epsilon\to0}\left|\left|\frac{\hat{U}(R,\bs{v},\bs{a},b+\epsilon)-\hat{U}(R,\bs{a},\bs{v},b)}{\epsilon}f-(-i)\hat{U}(I,\bs{v},\bs{a},b)\hat{H}f\right|\right|\qquad&&\quad\nonumber\\
=\lim_{\epsilon\to0}\left|\left|\frac{\hat{U}(I,\bs{0},\bs{0},\epsilon)-\hat{I}}{\epsilon}f+i\hat{H}f\right|\right|&=&0.
\label{2.7}
\end{eqnarray}
The limit \eqref{2.7} does not exist for all $f\in L^2(\mathbb{R}^3\otimes\mathbb{R})$. Functions for which it does exist form a dense subspace of 
$L^2(\mathbb{R}^3\otimes\mathbb{R})$ on which the operator $\hat{H}$ is self-adjoint. This result can be verified either by using the explicit form of the representation 
\eqref{2.5} or by appealing to the general result that when the operators $\hat{U}$ furnish a continuous unitary representation of a Lie group, 
the limit analogous to \eqref{2.7} exists on a dense subspace of the Hilbert space for every one parameter subgroup of the group. The linear 
operator defined by the limit elements, called the generator of the corresponding one-parameter subgroup, is self-adjoint.  Moreover, 
there exists a common dense domain on which the generators of all one parameter subgroups of a finite dimensional Lie group 
are defined as self-adjoint operators~\cite{garding}. This is the property 
that ensures the Lie algebra representation \eqref{2.6}. 

Alternatively, by differentiating the matrix representation \eqref{2.4} with respect to the group parameters, we can obtain a representation of the Galilean 
Lie algebra by $5\times5$-matrices. Either from that representation or from the self-adjoint representation \eqref{2.6}, we can compute the 
commutation relations among basis elements that characterize the Galilei Lie algebra: 
\begin{equation}
\begin{array}{lll}
[\hat{J}_i,\hat{J}_j]=i\epsilon_{ijk}\hat{J}_k & [\hat{J}_i,\hat{K}_j]=i\epsilon_{ijk}\hat{K}_k &[\hat{J}_i,\hat{P}_j]=i\epsilon_{ijk}\hat{P}_k\\[6pt]
[\hat{K}_i,\hat{K}_j]=0 & [\hat{P}_i,\hat{P}_j]=0 & [\hat{K}_i,\hat{P}_j]=0\\[6pt]
[\hat{J}_i,\hat{H}]=0 & [\hat{P}_i,\hat{H}]=0 & [\hat{K}_i,\hat{H}]=i\hat{P}_i.
\end{array}\label{2.8}
\end{equation}

The representation \eqref{2.5} is not physically very relevant. The physically important representations of the Galilei group are the unitary \emph{projective} 
representations. Recall that a representation of a Lie group $G$ is projective if, for every $g_1,g_2\in G$, the group operators fulfill the 
composition rule 
\begin{equation}
\hat{U}(g_2)\hat{U}(g_1)=e^{i\omega(g_2,g_1)}\hat{U}(g_2g_1)\label{2.9},
\end{equation}
where $\omega :\ G\times G\to\mathbb{R}$ is a real function. A projective representation is a special case of a two-cocycle representation in which the function 
$\omega$ depends also on the vector on which the operator identity \eqref{2.9} acts. This is the case for the representations of the Galilean line group that we will construct in 
section \ref{sec4}.

The function $\omega$ of \eqref{2.9} is determined, though not uniquely, by the structure of the group. The main restriction on $\omega$ is the associativity of the composition rule \eqref{2.9}, which gives 
\begin{equation}
\omega(g_3,g_2g_1)+ \omega(g_2,g_1)=\omega(g_3g_2,g_1)+\omega(g_3,g_2).\label{2.9b}
\end{equation}

The function $\omega$ has two distinct origins, one topological and the other 
algebraic. In the topological case, a non-trivial phase factor in \eqref{2.9} corresponds to a non-trivial first homotopy group (i.e., a group manifold that is not simply connected). 
Such is the case for both the Galilei and Poincar\'e groups, essentially 
owing to the non-simply connectedness of the rotation subgroup. For both groups, the structure of the rotation group leads to a phase factor equal to $\pm$ in \eqref{2.9}. For either group, 
this sign ambiguity can be removed by considering its universal covering group, obtained by essentially replacing the rotation group by $SU(2)$. 

In the algebraic case, a non-trivial phase factor in \eqref{2.9} corresponds to a central extension of the group or, equivalently, a central extension of its Lie algebra. While there are no such extensions of the Poincar\'e algebra, 
there exist central extensions of the Galilei algebra. In the basis used in  \eqref{2.8}, this extension corresponds to a change of the commutation relations between Galilei boosts 
and momenta from $[\hat{K}_i,\hat{P}_j]=0$  to 
\begin{equation}
[\hat{K}_i,\hat{P}_j]=i\delta_{ij}\hat{M}.\label{2.10}
\end{equation}
Here, $\hat{M}$ is an operator that commutes with the entire operator Lie algebra of \eqref{2.8}. It extends the Galilean algebra from a 10 dimensional algebra to an 11 dimensional algebra. 
Note that $\hat{M}$ is both a basis element of this extended Lie algebra and a Casimir operator of its enveloping algebra. 
The function $\omega:\ G\times G\to\mathbb{R}$ for the projective representations 
of the Galilei group corresponding to the central extension \eqref{2.10} is~\cite{bargmann}:
\begin{equation}
\omega(g_2,g_1)= \frac{1}{2} m (\bs{a}_2 \cdot R_2\bs{v}_1 - \bs{v}_2\cdot R_2\bs{a}_1 + b_1\bs{v}_2\cdot R_2\bs{v}_1)\label{2.11},
\end{equation}
where $g_i=(R_i,\bs{v}_i,\bs{a}_i,b_i)$ for $i=1,2$. The $m$ in \eqref{2.11} is a real valued parameter that arises as the eigenvalue of the mass operator $\hat{M}$. 
Since $\hat{M}$ is a central element of the associative algebra of the centrally extended Galilei algebra, 
in an irreducible representation, $\hat{M}=m\hat{I}$ for a fixed real number $m$. 

In order to construct the central extension of the Galilei group that corresponds to the extension of the algebra \eqref{2.10}, let 
\begin{equation}
\tilde{g}:=(\varphi,g), \qquad \varphi\in\mathbb{R},\ g\in{\cal G}\label{2.12}
\end{equation} 
and define an associative composition law as follows: 
\begin{equation}
\tilde{g}_2\tilde{g}_1=(\varphi_2+\varphi_1+\frac{1}{m}\omega(g_2,g_1), g_2g_1)\label{2.13},
\end{equation}
where $\omega(g_1,g_2)$ is given by \eqref{2.11} and the composition $g_2g_1$ is 
that given by \eqref{2.1}. It is straightforward to verify that the 
set of elements $\tilde{\cal G}=\{\tilde{g}\}$ is a group under \eqref{2.13}. 
It is clear that $\{(\varphi,e)\}$ is a 
central subgroup of $\tilde{\cal G}$.  
Furthermore, a unitary projective representation of the Galilei group defined by \eqref{2.11} is equivalent to a 
true representation of the centrally extended group. To see this, 
define 
\begin{equation}
\hat{U}(\tilde{g})=e^{im\varphi}\hat{U}(g)\label{2.14},
\end{equation} 
where $\hat{U}(g)$ are the unitary operators furnishing a projective representation of the Galilei group. It then follows from \eqref{2.9}, \eqref{2.11} and \eqref{2.13} 
that 
\begin{equation}
\hat{U}(\tilde{g}_2)\hat{U}(\tilde{g}_1)=\hat{U}(\tilde{g}_2\tilde{g}_1).\label{2.15}
\end{equation}
Furthermore, we recognize the mass operator $\hat{M}$ that centrally extends the Galilei Lie algebra \eqref{2.10} as the generator of the one parameter central subgroup 
$\hat{U}(\varphi,e)$ of the representation \eqref{2.14}. 

It is known that all of the unitary irreducible projective representations of the Galilei group, or equivalently, unitary irreducible true representations of its central extension, 
 can be constructed by Wigner's method of induced representations.  Since these matters are well known, we give only a very brief 
 summary of some important results that we will rely on in later sections. Detailed discussions and proofs of these results can be found, for instance, 
 in the comprehensive review article  \cite{levy}.  
 
In a unitary irreducible representation,  the three Casimir operators of the centrally extended Galilei algebra are all  proportional to the identity and define the 
representation: 
\begin{subequations}
\label{2.16}
\begin{eqnarray}
\hat{C}_1&=&\hat{M}=m\hat{I},\quad m\in\mathbb{R}\\
\hat{C}_2&=&\left(\hat{\bs{J}}^2-\frac{1}{\hat{M}}\hat{\bs{K}}\times\hat{\bs{P}}\right)^2=j(j+1)\hat{I},\quad j=0,1/2,1,3/2,\cdots\\
\hat{C}_3&=&\hat{H}-\frac{1}{2\hat{M}}\hat{\bs{P}}^2=w\hat{I},\quad w\in\mathbb{R}\label{2.16c}
\end{eqnarray} 
\end{subequations}
Therefore, each unitary irreducible representation of the extended Galilei group 
is characterized by three numbers $(m,j,w)$ that have interpretation as mass, spin and internal energy quantum numbers. 

The Hilbert space itself can be obtained as the space of $L^2$-functions defined on the Cartesian product of the 
spectra of a complete system of commuting operators (CSCO). A CSCO can be chosen to consist of, along with 
the Casimir operators \eqref{2.16}, the generators of the largest Abelian invariant subgroup ${\cal C}$ and the largest set of 
commuting generators of the ``little group". Recall that the little group is the largest subgroup of the factor group  ${\cal G}/{\cal C}$ that leaves a given 
(generalized) eigenspace of ${\cal C}$ invariant. For the centrally extended Galilei group, ${\cal C}$ is the subgroup of spacetime and central 
translations $(\varphi, I,\bs{0},\bs{a},b)$ and the little group is isomorphic to the rotation group.  In order to avoid nonessential complications, 
we will only consider the spinless case, i.e., only the one-dimensional trivial representation of the little group, in the remainder of this paper. 

Aside from the mass operator, the generators of the Abelian subgroup ${\cal C}=\{(\varphi, I,\bs{0},\bs{a},b)\}$ all have continuous spectra. A unitary 
irreducible representation of $\tilde{\cal G}$ can be completely defined by the action of operators $\hat{U}(\tilde{g})$ on the generalized eigenvectors 
of ${\cal C}$. Denoting these eigenvectors by $\left.\right|\bs{q}, E, [m,j=0,w]\rangle$, we then have
\begin{equation}
\hat{U}(\varphi,I,\bs{0},\bs{a},b)\left.\right|\bs{q}, E, [m,j=0,w]\rangle=e^{im(\varphi+\bs{q}\cdot\bs{a})-ibE}\left.\right|\bs{q}, E, [m,j=0,w]\rangle\label{2.17}.
\end{equation}
Consequently, for the generators of $\hat{U}(\varphi,I,\bs{0},\bs{a},b)$, 
\begin{subequations}
\label{2.18}
\begin{eqnarray}
\hat{M}\left.\right|\bs{q}, E, [m,j=0,w]\rangle&=&m\left.\right|\bs{q}, E, [m,j=0,w]\rangle\\
\hat{H}\left.\right|\bs{q}, E, [m,j=0,w]\rangle&=&E\left.\right|\bs{q}, E, [m,j=0,w]\rangle\\
\hat{\bs{P}}\left.\right|\bs{q}, E, [m,j=0,w]\rangle&=&m\bs{q}\left.\right|\bs{q}, E, [m,j=0,w]\rangle\label{2.18c}.
\end{eqnarray}
\end{subequations}
The transformation of $\left.\right|\bs{q},E, [m,j=0,w]\rangle$  under a general element of $\tilde{\cal G}$ is given by 
\begin{equation}
\hat{U}(\tilde{g})\left.\right|\bs{q},E,[m,j=0,w]\rangle=e^{im(\varphi+\bs{q}'\cdot\bs{a}-\frac{1}{2}\bs{v}\cdot\bs{a})-bE'}\left.\right|\bs{q}',E',[m,j=0,w]\rangle\label{2.19},
\end{equation}
where 
\begin{eqnarray}
\bs{q}'&=&R\bs{q}+\bs{v}\nonumber\\
E'&=&E+mR\bs{q}\cdot\bs{v}+\frac{1}{2}m\bs{v}^2\label{2.18b}.
\end{eqnarray}
It follows that 
\begin{equation}
E'-\frac{1}{2}m{\bs{q}'}^2=E-\frac{1}{2}m\bs{q}^2.\label{2.19b}
\end{equation} 
That is to say, the difference between the total energy and kinetic energy, generally understood as the internal energy, is a Galilean invariant; this result is as anticipated 
from \eqref{2.16} as well as classical Newtonian physics.

 For notational simplicity, we may  suppress all of the
 quantum numbers pertaining to the Casimir operators in the eigenvectors when dealing with a particular irreducible representation.
Furthermore, in view of \eqref{2.19b} or \eqref{2.16}, we note that the generalized eigenvalue $E$ of the Hamiltonian 
 $\hat{H}$ is not an independent variable in an irreducible representation but completely determined by $m$, $\bs{q}$ and $w$. 
 In light of these considerations, we may suppress the energy variable $E$ as well and 
 denote the basis vectors simply by $\left.\right|\bs{q}\rangle$. Therewith, the transformation formula \eqref{2.19} 
 under a general element of $\tilde{\cal G}$ can be rewritten as
\begin{equation}
\hat{U}(\tilde{g})\left.\right|\bs{q}\rangle=e^{im(\varphi+\bs{q}'\cdot\bs{a}-\frac{1}{2}\bs{v}\cdot\bs{a})-iE'b}\left.\right|\bs{q}'\rangle.\label{2.20}
\end{equation}

In the next section, we will consider the Galilean line group as a candidate for transformations 
amongst inertial as well as accelerating reference frames.  We will show the interesting property that there are no central extensions of the line group that contain a given central extension 
of the Galilei group. This poses a problem for defining the concept of inertial mass for a quantum physical system described by a unitary representation of the Galilean line group, and in particular for 
formulating a quantum mechanical principle of equivalence. Nevertheless, in the absence of rotational accelerations, 
we will show that there do exist certain \emph{non-central} extensions of the Galilean line group containing the central extensions of the Galilei
group. These extensions allow the definition of an inertial mass and lead to the equivalence principle. In this light, it is the (non-centrally) extended Galilean line 
group that we propose as physically relevant for formulating quantum mechanics in non-inertial reference frames. 

\section{The Galilean line group}\label{sec3}
As outlined in the preceding section, unitary projective representations of the Galilei group provide an example of the mathematical 
synthesis of the two key physical principles of quantum theory:  superposition and relativity. As noted above, in this paper we  deal with 
non-relativistic (Galilean) quantum mechanics. Relativistic (Lorentzian) quantum mechanics 
would require unitary projective representations of the inhomogeneous Lorentz group.  

In either Galilean or Lorentzian  formulations of quantum mechanics, the theory is restricted to inertial reference frames. In order to extend the formulation to include non-inertial reference frames, 
it is natural to examine whether transformations can be defined amongst accelerating reference frames, both linear and rotational, so as to form a group 
and then construct this group's unitary, possibly projective, representations. Clearly, we must require that these acceleration transformations contain 
 either Galilean transformations or Poincar\'e transformations as a subgroup, depending on whether we seek an extension of Galilean or 
of Lorentzian  quantum mechanics.  Our focus in this paper is the former. That is, we wish to incorporate time dependent boosts and rotations, 
rather than only the constant $R$ and $\bs{v}$ of the Galilei group, while keeping the group structure much the same as that of the Galilei group.  The 
representations of this acceleration group will give rise to a fictitious potential energy term in the Hamiltonian when the Hamiltonian is transformed to a non-intertial 
reference frame.  This fictitious potential energy term has a natural interpretation as a gravitational potential energy. While such a gravitational field can have an arbitrary 
(analytic) time dependence, it must have a zero spacial gradient. This is the main limitation of our construction. 

In order to further generalize the theory to accommodate gravitational fields with non-trivial spacial dependence, we must try to introduce a notion of ``locality'' to the 
acceleration group. The simplest such approach appears to be letting the domain of analytic functions into the Galilei group be the 
entire spacetime manifold $\mathbb{R}\otimes\mathbb{R}^3$, rather than the only the time axis $\mathbb{R}$. Such an extension would be in the spirit 
of Einstein's introduction of local coordinate systems at every spacetime point. In Appendix \ref{sec3.5}
we will show that the set of analytic functions from $\mathbb{R}\otimes\mathbb{R}^3$ to $\cal G$ form 
a semigroup under a composition law analogous to that of the Galilei group. We can view this semigroup as a generalized Galilei group in which the rotations, boosts and  translations 
 depend on position in space-time. At this point, it remains an open problem to determine if an inverse exists 
for every element and, should that be the case, to construct the representations of the resulting group.

\subsection{Acceleration transformations}\label{sec3.1}

The boost and space translation terms of \eqref{2.3} can be combined into a time-dependent space translation $\bs{a}(t)=\bs{v}t+\bs{a}$.  Thus \eqref{2.3} 
can be more compactly written as
\begin{eqnarray}
\bs{x}'&=&R\bs{x}+\bs{a}(t)\nonumber\\
t'&=&t+b\label{3.1.1}.
\end{eqnarray}
 
This structure motivates us to consider arbitrary analytic functions for $\bs{a}$ and $R$. To that end, consider the set of transformations defined by
\begin{equation}
(R,\bs{a},b):\quad \left(\begin{array}{c}
\bs{x}\\
t\end{array}
\right)\to 
\left(\begin{array}{c}
\bs{x}'\\
t'\end{array}
\right)=
\left(\begin{array}{c}
R(t)\bs{x}+\bs{a}(t)\\
t+b\end{array}
\right)\label{3.1.2}.
\end{equation}
The $R$ and $\bs{a}$ are now functions of time, taken to be (real) analytic.  With \eqref{3.1.1} or \eqref{3.1.2},  the boost element is now defined as the first time derivative of the spatial translation, 
rather than as a separate parameter. To recover the inertial case, we keep $R$ constant and let $\bs{a}(t)=\bs{a}_0+\bs{v}t$. 

In order to define a group composition rule on the set of functions $(R,\bs{a},b)$, consider the successive operation of two elements on a spacetime point: 
\begin{equation}
 \left(\begin{array}{c} \bs{x}'' \\ t''\end{array}\right) 
= \left(\begin{array}{c}R_2(t + b_1)R_1(t)\bs{x}+\bs{a}_2(t+b_1)+R_2(t+b_1)\bs{a}_1(t)\\
  t+b_1+b_2\end{array}\right). \label{3.1.3}
\end{equation} 
{\bf Closure:}\\
The closure of the set $\{(R,\bs{a},b)\}$ under the
 composition rule implied by \eqref{3.1.3} is not readily obvious because of the $b$'s appearing in the argument of the second element $(R_2,\bs{a}_2,b_2)$.  
However, with the use of the shift operator $\Lambda_b:\ f(t)\rightarrow \left(\Lambda_bf\right)(t)=f(t+b)$, we can rewrite \eqref{3.1.3} as
 \begin{eqnarray}
 \left(\begin{array}{c} \bs{x}'' \\ 
t''\end{array}\right)& =& \left(\begin{array}{c}(\Lambda_{b_1}R_2)
           (t )R_1(t)\bs{x}+(\Lambda_{b_1}R_2)(t)\bs{a}_1(t) + (\Lambda_{b_1}\bs{a}_2)(t)  \\
t+b_1+b_2\end{array}\right)\nonumber\\
&=&\left(\begin{array}{c}\left(\Lambda_{b_1}R_2R_1\right)(t)\bs{x}+\left(\Lambda_{b_1}R_2\bs{a}_1\right)(t) + (\Lambda_{b_1}\bs{a}_2)(t)  \\
t+b_1+b_2\end{array}\right),\label{3.1.4}
\end{eqnarray}
where the second equality involves the definition of the product of two (smooth) functions by way of the pointwise multiplication: 
$(f_2f_1)(t):=f_2(t)f_1(t)$.  From \eqref{3.1.4}, we extract a composition rule for the set of functions $(R,\bs{a},b)$: 
\begin{equation}
(R_2, \bs{a}_2, b_2)(R_1, \bs{a}_1, b_1) = (\Lambda_{b_1} R_2 R_1 , \ \Lambda_{b_1} R_2\bs{a}_1 + \Lambda_{b_1} \bs{a}_2 ,\  b_1 + b_2)\label{3.1.5}.
\end{equation}
Since analytic functions form an algebra under pointwise multiplication and since this algebra remains invariant under the 
shift operator $\Lambda_b$ for all $b\in\mathbb{R}$, it follows from the composition rule \eqref{2.1} that the set of functions $(R,\bs{a},b)$ is closed 
under the composition rule  \eqref{3.1.5}.\\\\
{\bf Inverse:}\\ 
Under \eqref{3.1.5}, each element $(R,\bs{a},b)$ also has an inverse: 
\begin{equation}
(R, \bs{a}, b)\inv = (\Lambda_{-b} R\inv, -\Lambda _{-b}(R\inv \bs{a}) , -b).\label{3.1.6}
\end{equation}
{\bf Associativity:}\\
 Composing three elements shows that the composition rule \eqref{3.1.5} is
associative if, for all $b_1$ and $b_2$, the following hold: 
\begin{subequations}
\label{3.1.7}
\begin{eqnarray}
\Lambda_{b_2} \Lambda_{b_1} &= &\Lambda_{b_2+ b_1} \\
\Lambda_{b}(R\bs{a}) &= &(\Lambda_b R)(\Lambda_b \bs{a} )\\
\Lambda_b (R_1 R_2) &= &(\Lambda_b R_1)(\Lambda_b R_2)\label{3.1.7c}.
\end{eqnarray}
\end{subequations}
Each of these properties follows from our definition of $\Lambda_b$ as the shift operator. Thus, we conclude that the 
set of analytic functions $(R,\bs{a},b)$ is a group under \eqref{3.1.5}, which we will refer to as the Galilean line group. 
For the sake of the economy of notation, we will denote the Galilean line group also by ${\cal G}$. 

Note that ${\cal R}:=\{(R,\bs{0},0)\}$,  ${\cal A}:=\{(I,\bs{a},0)\}$ and ${\cal B}:=\{(I,0,b)\}$ are all 
subgroups of the Galilean line group.  The subgroup ${\cal E}(3):=\{(R,\bs{a},0)\}$, which is the line group 
of the three dimensional Euclidean group,  is the semidirect product of 
${\cal A}$ and ${\cal R}$: 
\begin{equation}
\mathcal{E}(3)=\mathcal{A}\rtimes\mathcal{R}.\label{3.1.8}
\end{equation}
Furthermore, from \eqref{3.1.5} we note that each $(I,\bs{0},b)\in{\cal B}$ induces an automorphism on the subgroup ${\cal E}(3)$, given by $\Lambda_b$. 
Therefore, ${\cal G}$ is the semidirect product of ${\cal E}(3)$ and ${\cal B}$: 
\begin{equation}
\mathcal{G}  = {\mathcal E}(3)\rtimes_{\Lambda}\mathcal{B}= ( \mathcal{A} \rtimes \mathcal{R} ) \rtimes  _{\Lambda} \mathcal{B}.\label{3.1.10}
\end{equation}
That is, the Galilean line group is the semidirect product of the time translation subgroup and the line group of the three dimensional Euclidean group.
(It is worth noting that our use of the external semidirect product differs slightly from the canonical definition.
 Looking at the subgroup of space and time translations, we have $(a_2, b_2)(a_1, b_1) = (\Lambda_{b_1}a_2 + a_1, b_2 + b_1)$, whereas a connonical external semi-direct product would give $(a_2, b_2)(a_1, b_1) = (a_2 + \Lambda_{b_2}a_1, b_2 + b_1)$.     
Had we chosen our notation so that multiple transformations composed from left to right instead of right to left, then the stucture of the Galilei Galilean Line Group would look exactly like that of a semidirect product. That is to say, our semidirect product definition is the dual of the more standard one.)

\subsection{Galilean line group  as an infinite parameter topological group}\label{sec3.2}
Recall that an analytic function has a converging Taylor series given by 
\begin{equation}
f(t)=\sum_n\frac{1}{n!}f^{(n)}t^n, \quad f^{(n)}:=\left.\frac{d^nf}{dt^n}\right|_{t=0},
\end{equation}
where $f$ is uniquely determined by the coefficients $f^{(n)}$. Therefore, instead of the analytic functions $R$ and $\bs{a}$ that determine ${\cal E}(3)$ of \eqref{3.1.8}, we may consider the Taylor coefficients of $R$ and $\bs{a}$. 
This allows us to consider  the Galilean line group ${\cal G}$ as an infinite dimensional topological group defined 
by the parameters $(\{\bs{R}^{(0)},\bs{R}^{(1)},\bs{R}^{(2)}...\},\{\bs{a}^{(0)},\bs{a}^{(1)},\bs{a}^{(2)}...\}, b)$, rather than analytic functions 
$R$ and $\bs{a}$ from $\mathbb{R}$ to the Euclidean group.  Here, $\bs{a}^{(a)}$ and $\bs{R}^{(n)}$ collectively denotes the three Taylor coefficients 
\begin{equation}
a_i^{(n)}=\left.\frac{d^na_i}{dt^n}\right|_{t=0}\quad\mathrm{and}\quad
\left.R_{i}^{(n)}=\frac{d\theta}{dt}\frac{dR_i(\theta(t))}{d\theta}\right|_{t=0}
\end{equation} 
respectively, where $i=1,2,3$. $R_i(\theta)$ denotes a rotation about the $x_i$-axis by angle $\theta(t)$.  

The automorphism $\Lambda_b$ can also be explicitly defined in terms of its action on the Euclidean line group elements 
$(\{\bs{R}^{(0)},\bs{R}^{(1)},\bs{R}^{(2)}\},\cdots, \{\bs{a}^{(0)},\bs{a}^{(1)},\bs{a}^{(2)},\cdots\})$.  To that end, we simply consider the Taylor expansion of the functions 
$\left(\Lambda_b\bs{a}\right)(t)=\bs{a}(t+b)$ and $\left(\Lambda_bR\right)(t)=R(t+b)$ around $t=0$ and gather all coefficients for each $t^N$. 
For $\Lambda_b{\bs{a}}$, we obtain
\begin{eqnarray}
\left(\Lambda_b\bs{a}\right)(t) &=&\bs{a}(t+b)\nonumber\\
&=& \sum_{n=0}^{\infty}\frac{1}{n!} \bs{a}^{(n)} (t+b)^n\nonumber\\
& =& \sum_{n=0}^{\infty}\frac{1}{n!} \bs{a}^{(n)} \left( \sum_{k=0}^{n} \left( \begin{array}{c}n\\k\end{array}\right) b^{n-k}t^k \right).\label{3.2.1}
\end{eqnarray}
For a given $N$, the coefficient of $t^N$ comes from those terms with $k=N$, which occurs only for $n\geq N$. Thus the $t^N$ terms are
\begin{equation}
\sum_{n=N}^\infty \frac1{n!}\bs{a}^{(n)}  \left( \begin{array}{c}n\\N\end{array}\right)b^{n-N}t^N.\label{3.2.2}
\end{equation}
Replacing $n$ with $m$, $N$ with $n$, and summing over all powers of $t$ gives
\begin{eqnarray}
 \sum_{n=0}^{\infty}\frac{1}{n!} \bs{a}^{(n)} (t+b)^n  
&  =&\sum_{n=0}^\infty \sum_{m=n}^\infty  \frac{1}{m!}    \left( \begin{array}{c}m\\n\end{array}\right)   \bs{a}^{(m)} b^{m-n} t^n \nonumber\\
& =&\sum_{n=0}^\infty \frac{1}{n!} \left( \sum_{m=n}^\infty  \frac{1}{(m-n)!} \bs{a}^{(m)} b^{m-n}\right) t^n .\label{3.2.3}
\end{eqnarray}
From \eqref{3.2.1} and \eqref{3.2.3}, we then have 
\begin{equation}
\Lambda_b\bs{a}(t)=\sum_{n=0}^\infty \frac{1}{n!} \left( \sum_{m=n}^\infty  \frac{1}{(m-n)!} \bs{a}^{(m)} b^{m-n}\right) t^n \label{3.2.4},
\end{equation}
i.e. $(\Lambda_b\bs{a})^{(n)} = \sum_{m=n}^\infty \frac{1}{(m-n)!}\bs{a}^{(m)} b^{m-n}$. Therefore, the action of $\Lambda_b$ on $\bs{a}=\{\bs{a}^{(0)},\bs{a}^{(1)},\bs{a}^{(2)},\cdots\}$ is given by
\begin{eqnarray}
\Lambda_b:\ \{\bs{a}^{(0)},\bs{a}^{(1)},\bs{a}^{(2)},\cdots\}\to\left\{\sum_{m=0}^\infty \frac{1}{(m)!}\bs{a}^{(m)} b^{m}, \sum_{m=1}^\infty \frac{1}{(m-1)!}\bs{a}^{(m)} b^{m-1}, \sum_{m=2}^\infty \frac{1}{(m-2)!}\bs{a}^{(m)} b^{m-2},\cdots\right\}.\nonumber\\\label{3.2.5}
\end{eqnarray}
A similar calculation can be carried out to determine the action of $\Lambda_b$ on $\{\bs{R}^{(0)},\bs{R}^{(1)},\bs{R}^{(2)},\cdots\}$.

However, due to the cumbersome nature of this notation, in general we will parametrize the Galilean line group by analytic functions, writing the elements of the group as $(R,\bs{a},b)$, 
 where $R$ and $\bs{a}$ acquire the values $R(t)$ and $\bs{a}(t)$, respectively, when the group element acts on a point in space time $(\bs{x},t)$.  
Obviously, $\bs{a}(t)=\bs{a}^{(0)}+\bs{a}^{(1)}t+\frac{1}{2!}\bs{a}^{(2)}t^2+\cdots$ and $R(\bs{\theta}(t))=\bs{R}^{(0)}+\bs{R}^{(1)}t+\frac{1}{2!}\bs{R}^{(2)}t^2+\cdots$, 
where $t^n$ can be thought of as a place holder for $\bs{a}^{(n)}$ and $\bs{R}^{(n)}$.   
\\

\subsection{Generators and commutation relations}\label{sec3.3}
With the above characterization of $\cal G$ as an infinite parameter topological group, we can define its algebra as the linear span of tangent 
vectors of all one parameter subgroups, evaluated at the group identity. A basis for a representation of this algebra on a Hilbert space can be obtained by 
differentiating one parameter operator subgroups for each $\hat{U}\left(R_i^{(n)}\right)$, $\hat{U}\left({a}_i^{(n)}\right)$ and 
$\hat{U}\left(b\right)$ and evaluating the derivatives at the identity. 
Thus, for example, we define the generators $\hat{K}^{(n)}_i$ associated with $a^{(n)}_i$ as usual:
\begin{equation}
\hat{K}^{(n)}_i = i\left. \frac{d\hat{U}(I,{a}^{(n)}_i,0)}{d a^{(n)}_i}\right|_{\bs{a} = \bs{0}} . \label{3.3.1}
\end{equation}
Similarly,  the Hamiltonian and the generators for the rotation parameters can be defined by 
\begin{eqnarray}
\hat{H}&=& i \left. \frac{d\hat{U}(I,\bs{0},b)}{db}\right|_{b= 0}\label{3.3.2}\\\nonumber\\
\hat{J}^{(n)}_i&=&i\left.\frac{d\hat{U}(R(\theta_i^{(n)}),\bs{0},0)}{d\theta^{(n)}_i}\right|_{\bs{\theta}=0} . \label{3.3.3}
\end{eqnarray}

A concrete realization of these operators can be obtained, as done in \eqref{2.5} and \eqref{2.6}  for the Galilei group, by constructing a unitary representation of the Galilean line group 
in the $L^2$-function space on $\mathbb{R}\otimes\mathbb{R}^3$. To that end, let 
\begin{eqnarray}
\left(\hat{U}(R,\bs{a},b)f\right)(\bs{x},t)&=&f\left((R,\bs{a},b)^{-1}(\bs{x},t)\right)\nonumber\\
&=&f\left(\Lambda_{-b}R^{-1}(t)\bs{x}-\left(\Lambda_{-b}\left(R^{-1}\bs{a}\right)\right)(t), t-b\right).
\label{3.3.4}
\end{eqnarray}
Now, using definition \eqref{3.3.1}, 
\begin{eqnarray}
\left(\hat{K}^{(n)}_i f\right)(\bs{x}, t)& =&i \left. \frac{d\hat{U}}{d{a}^{(n)}_i}f(x,t) \right| _{\bs{a}, b = 0} \nonumber\\
&=&i\left. \lim_{\varepsilon \rightarrow 0}\frac{ \hat{U}({a}^{(n)}_i+\varepsilon) - 
\hat{U}(a^{(n)}_i)}{\varepsilon}f(x, t)\right| _{\bs{a}=0}\nonumber\\
&=&i\left. \lim_{\varepsilon\rightarrow 0} \frac{f(\bs{x} -\hat{i}\frac{1}{n!}({a}_i^{(n)} +\varepsilon) t^n, t) - 
f(\bs{x} - \hat{i}\frac{1}{n!}a_i^{(n)}t^n, t)}{\varepsilon}\right| _{\bs{a}= 0},\label{3.3.5}
\end{eqnarray}
where $\hat{i}$ is the unit vector in the $i^{\rm th}$-direction. Substituting $\delta = -\frac{1}{n!}t^n \varepsilon$ gives
\begin{equation}
\lim_{\delta \rightarrow 0}\frac{ -\frac{1}{n!} t^n f(\bs{x} - \delta \hat{i}, t) - f(\bs{x}, t)} {\delta} = -\frac{t^n}{n!} \frac{df}{dx_i}.\label{3.3.6}
\end{equation}
So we have
\begin{equation}
\left(\hat{K}^{(n)}_if\right)(\bs{x},t) = -i\frac{t^n}{n!}\frac{d}{dx_i}f(\bs{x},t)=\frac{t^n}{n!}\hat{K}^{(0)}_if(\bs{x},t).\label{3.3.7}
\end{equation} 
A similar derivation gives 
\begin{eqnarray}
\left(\hat{H}f\right)(\bs{x},t) &=& -i\frac{d}{dt}f(\bs{x},t)\label{3.3.8}\\
\left(\hat{J}^{(n)}_if\right)(\bs{x},t)&=&-i\frac{t^n}{n!}\epsilon_{ijk}x_j\frac{d}{dx_k}f(\bs{x},t)=\frac{t^n}{n!}\hat{J}_if(\bs{x},t)\label{3.3.9}
\end{eqnarray}
Note that all infinity of operators are in fact obtained from the finite set $\hat{H}$, $\hat{\bs{K}}^{(0)}$ and $\hat{\bs{J}}^{(0)}$. 
Further, since $\hat{\bs{K}}^{(0)}$ and $\hat{\bs{K}}^{(1)}$ are the generators of pure space translations $\bs{a}^{(0)}$ and Galilean boosts $\bs{a}^{(1)}$ 
respectively, they coincide with the momentum operators $\hat{\bs{P}}$ and boost operators $\hat{\bs{K}}$ of the Galilean Lie algebra \eqref{2.8}.

Just as was the case for the Galilean algebra, these operators are generally not 
defined on the whole of $L^2\left(\mathbb{R}\otimes{\mathbb{R}}^3\right)$. However, from the explicit calculation leading  to \eqref{3.3.7}, we note that they are densely defined in 
$L^2\left(\mathbb{R}\otimes{\mathbb{R}}^3\right)$ and, if  the $\hat{U}(R,\bs{a},b)$ are unitary (with respect to, say, the inner product defined by the usual Lesbegue measure), 
then \eqref{3.3.7}-\eqref{3.3.9} are self-adjoint on this domain. An invariant common 
dense domain for all the generators \eqref{3.3.7}-\eqref{3.3.9} is also likely to exist in view of the fact that all generators are in fact obtained 
from the finite set $\hat{H}$, $\hat{\bs{K}}$ and $\hat{\bs{J}}$. Should this indeed be the case (a subtle mathematical issue 
that we will not investigate in this paper), then the ${\hat{\bs{K}}}^{(n)}$, ${\hat{\bs{J}}}^{(n)}$ and $\hat{H}$ 
will span an operator algebra. The commutation relations amongst these basis elements can also be readily computed 
from the explicit form given by \eqref{3.3.7}-\eqref{3.3.9}: 
\begin{subequations}
\label{3.3.10}
\begin{eqnarray}
[\hat{K}^{(n)}_i ,\hat{K}^{(m)}_j] & =& 0\label{3.3.10a}\\ [6pt]
[\hat{H}, \hat{K}^{(n)}_i] &=& -i\hat{K}^{(n-1)}_i\ \ \text{for}\ n\geq1;\quad [\hat{H}, \hat{K}^{(0)}_i]= 0\\[6pt]
[\hat{J}^{(n)}_i,\hat{K}^{(m)}_j] &=&i\frac{t^n}{n!}\epsilon_{ijk}\hat{K}^{(m)}_k=i\frac{t^m}{m!}\epsilon_{ijk}\hat{K}^{(n)}_k\\[6pt]
[\hat{J}^{(n)}_i,\hat{J}^{(m)}_j]&=&i\frac{t^n}{n!}\epsilon_{ijk}\hat{J}^{(m)}_k=i\frac{t^m}{m!}\epsilon_{ijk}\hat{J}^{(n)}_k\\[6pt]
[\hat{H},\hat{J}^{(n)}_i]&=&i\hat{J}^{(n-1)}_i\ \ \text{for}\ n\geq1;\quad [\hat{H}, \hat{J}^{(0)}_i]= 0.\label{3.3.10e}
\end{eqnarray}
\end{subequations}

\subsection{Extensions of the Galilean line group}\label{sec3.4}
\subsubsection{Central Extensions of Lie Algebra}
As noted above in Section \ref{sec2}, physically relevant representations of the Galilei group are projective representations, as it is only in 
these representations that a mass operator and a position operator canonically conjugated to a momentum 
operator can be defined, see \eqref{2.10}. Recall also that a unitary irreducible projective representation of the Galieli group is 
equivalent to a unitary irreducible true representation of the centrally extended Galilei group, furnished by \eqref{2.11} and \eqref{2.13}. 

If we expect that a physically meaningful representation of the Galilean line group should always contain a physically 
meaningful representation of the Galilei group, then we must construct a suitable extension of the former that contains the central 
extension \eqref{2.13} of the latter. The simplest choice is to construct a central extension of the Galilean line group, or equivalently 
a central extension of the  algebra \eqref{3.3.10}, that contains \eqref{2.13} as a subgroup. 
In particular, the commutation relations of the centrally extended algebra \eqref{3.3.10} must contain \eqref{2.10}. 
\emph{However, although the Galilei group is naturally contained in the Galilean line group, 
the rather surprising conclusion of the simple calculation below is that there exist no central extensions of the Galilean line group that 
contain a given central extension of the Galilei group.}

Since the crucial property we seek is the embedding of \eqref{2.10} amongst the commutation relations of the extended algebra \eqref{3.3.10}, 
we need only consider the algebra of the space and time translation subgroup of the Galilean line group.  
Suppose we construct a central extension of this algebra by letting 
\begin{equation}
[\hat{K}^{(n)}_i ,\hat{K}^{(m)}_j ]  = \alpha_{mn} \delta_{ij}   \label{3.4.1}
\end{equation}
for some (central) scalar $\alpha_{mn}$.

If we take at least $\hat{\bs{K}}^{(0)}, \hat{\bs{K}}^{(1)}, \hat{\bs{K}}^{(2)}$ to be nonzero -- so that this algebra properly contains that of the Galilei case - 
then by the Jacobi identity for $\hat{\bs{K}}^{(0)}$, $\hat{\bs{K}}^{(2)}$ and $\hat{H}$,  
\begin{subequations}
\label{3.4.2}
\begin{eqnarray}
[\hat{H}, [\hat{K}^{(0)}_i, \hat{K}^{(2)}_j]] + [\hat{K}^{(0)}_i, [\hat{K}^{(2)}_j, \hat{H}]] + [\hat{K}^{(2)}_j, [\hat{K}^{(0)}_i, \hat{H}]]&=&0\\
-i[\hat{K}^{(0)}_i, \hat{K}^{(1)}_j]&=&0\label{3.4.2b}\\
\alpha_{01}&=&0.\label{3.4.2c}
\end{eqnarray}
\end{subequations}
Here we have used the commutation relation \eqref{3.3.10e} to obtain \eqref{3.4.2b}, followed by  \eqref{3.4.1} to infer \eqref{3.4.2c}. 
From equations \eqref{3.4.2} we see that the generators of constant spatial translations, $\hat{\bs{K}}^{(0)}:=\hat{\bs{P}}$, necessarily commute with those of boosts, $\hat{\bs{K}}^{(1)}:=\hat{\bs{K}}$,
whenever we incorporate reference frames that move with non-zero linear accelerations (the $\hat{\bs{K}}^{(2)}\not=0$ condition),
in contradiction to \eqref{2.10}. (In general, it can be seen from the same argument as above that for the subgroup consisting of functions $\bs{a}(t)$ with vanishing 
Taylor coefficients  $\bs{a}_n=0$ for all $n\geq N+1$, all $\alpha_{mn}$ of \eqref{3.4.1} vanish except for $\alpha_{N,N-1}$.) 
This shows that there are no central extensions of the algebra of the Galilean line group that contain central extensions of the Galilei group. 
Thus we will not be able to construct a projective representation of the Galilean 
line group containing any given projective representation of the Galilei group. 

\subsubsection{Other extensions}
The impossibility of embedding a central extension of the Galilei group in a central extension of the Galilean line group implies 
that the latter is not the relevant concept. To circumvent the difficulty, there are two obvious alternatives that we may consider: either 
we could look for a non-central extension of the line group that contains a given central extension of the Galilei group, or we could 
start with the centrally extended Galilei group and construct its line group, which would automatically solve the embedding problem. 
These two approaches lead to same result, but only if we restrict ourselves to contant rotations: the associativity of the composition law for the extended line group
and its proper reduction to a central extension of the Galilei group appear to be at odds with each other whenever we have time dependent rotations. 

To construct an extension of the Galilean line group, let $\varphi$ be a real-valued analytic function of time and define 
\begin{equation}
\tilde{g}:=\left(\varphi,g\right)\label{3.4.3}
\end{equation}
where $g=(R,\bs{a},b)$ is an element of the Galilean line group. From \eqref{3.1.3}, we expect that when we compose two elements of the form \eqref{3.4.3}, the function 
$\varphi$ will be evaluated at two different points in time, $t$ and $t+b$. Therefore, as in \eqref{3.1.4}, we expect the automorphism $\Lambda_b$ to act on functions $\varphi$ 
in any well-defined composition rule. (Incidentally, it is this fact that makes the extension \eqref{3.4.3}  non-central.) In view of this observation, we define a composition rule 
for \eqref{3.4.3} by 
\begin{equation}
\tilde{g}_2 \tilde {g}_1 = \left(\varphi_1 + \Lambda_{b_1}\varphi_2+ \frac{1}{m}\xi(g_2, g_1),\ g_2g_1\right)\label{3.4.4}
\end{equation}
where, just as in \eqref{2.13}, $\xi(g_2,g_1)$ denotes a real-valued function on the Galilean line group, 
$\xi:\ {\cal G}\otimes{\cal G}\to\mathbb{R}$ and $m$ is a real constant.  
In addition to the properties necessary to make the set of elements \eqref{3.4.3} a group 
under the composition rule \eqref{3.4.4}, we must also demand that \eqref{3.4.4} reduce to the composition rule \eqref{2.13} when $\varphi$ and $R$ are constant functions and 
$\bs{a}$ is of the form $\bs{a}(t)=\bs{v}t+\bs{a}_0$. 

If we require that $\xi$ be an analytic function, then the closure of the set of elements \eqref{3.4.3} under \eqref{3.4.4} follows from the closure of the algebra of analytic 
functions $\varphi$ under the shift operator $\Lambda_b$ and the closure of the Galilean line group.  The existence of an inverse for 
each $\left(\varphi,g\right)$ under \eqref{3.4.4} is also straightforward to verify. 

The associativity of \eqref{3.4.4}, i.e.
$\tilde{g}_1\left(\tilde{g}_2\tilde{g}_3\right)=\left(\tilde{g}_1\tilde{g}_2\right)\tilde{g}_3$,  requires that
\begin{eqnarray}
&&\left(\varphi_1 + \Lambda_{b_1}\varphi_2 + \Lambda_{b_1 + b_2}\varphi_3 + \frac{1}{m}\xi(g_2, g_1) + \frac{1}{m}\xi(g_3 , g_2 g_1), g_3(g_2g_1) \right)\nonumber\\
&&\quad= \left( \varphi_1 + \Lambda_{b_1}\varphi_2 + \Lambda_{b_1 + b_2}\varphi_3 + \frac{1}{m}\Lambda_{b_1}\xi(g_3, g_2) + \frac{1}{m}\xi(g_3 g_2 , g_1), (g_3g_2)g_1\right ).
\label{3.4.5}
\end{eqnarray}

This shows that there are no further restrictions on the part of the composition rule that involves the functions $\varphi$. However, the associativity requirement \eqref{3.4.5} 
does impose a restriction on the function $\xi$: 
\begin{equation}
\xi(g_2, g_1) + \xi(g_3 , g_2 g_1)=\Lambda_{b_1}\xi(g_3, g_2) + \xi(g_3 g_2 , g_1).\label{3.4.6}
\end{equation}
This differs from the corresponding equality \eqref{2.9b}  by the appearance of the shift operator $\Lambda_{b_1}$. 

The requirement that \eqref{3.4.4} be consistent with both \eqref{3.4.6} and \eqref{2.13} is satisfied when  rotations are time independent $\dot{R}=0$ and 
$\xi(g_2,g_1)$ is of the form 
\begin{equation}
\xi(g_2, g_1) =\frac{1}{2}m\Bigl( \Lambda_{b_1}\bs{a}_2\cdot  R_2 \dot{\bs{a}}_1 - \Lambda_{b_1} \dot{\bs{a}}_2 \cdot R_2 \bs{a}_1\Bigr),\ \ \text{for}\ \dot R=0\label{3.4.7}.
\end{equation}
To see that \eqref{3.4.7} does indeed reduce to \eqref{2.11}, we set $\bs{a}(t)=\bs{a}^{(0)}+\bs{v}t$, ($\bs{v}$ and $\bs{a}^{(0)}$ constant). 
Since the Galilei group is conventionally parametrized not by the time dependent 
space translation functions $\bs{a}(t)=\bs{a}^{(0)}+\bs{v}t$ but by their Taylor coefficients, to obtain \eqref{2.11} from \eqref{3.4.7} we must first set 
$\dot{\bs{a}}(0)=\bs{v}$ and then evaluate $\bs{a}(t)$ at $t=0$, $\bs{a}(0)=\bs{a}^{(0)}$.  
Note that care must be taken when evaluating functions acted upon by the 
shift operator, i.e., $\Lambda_b\bs{a}(0)=\bs{a}(b)$. Therewith, 
\begin{eqnarray}
\left.\xi((R_2,\bs{a}_2(t),b_2), (R_1,\bs{a}_1(t),b_1))\right|_{t=0} &= &\frac{1}{2}m\Bigl(\Lambda_{b_1}\bs{a}_2(0)\cdot  R_2 \bs{v}_1- \Lambda_{b_1}\bs{v}_2\cdot R_2 \bs{a}_1(0)\Bigr)\nonumber\\
&=&\frac{1}{2}m\Bigl((\bs{a}_2^{(0)}+\bs{v}_2(0+b))\cdot R_2\bs{v}_1-\bs{v}_2\cdot R_2\bs{a}_1^{(0)}\Bigr)\nonumber\\
&=&\frac{1}{2}m\Bigl(\bs{a}_2^{(0)}\cdot R_2\bs{v}_1-\bs{v}_2\cdot R_2\bs{a}_1^{(0)}+b\bs{v}_2\cdot R_2\bs{v}_1\Bigr)
\label{3.4.8}, 
\end{eqnarray}
which is exactly the Galilei group phase factor.

It is not clear if the requirement of time independent rotations assumed for \eqref{3.4.7} is indespnesible. The trouble is that when rotations have non-trivial time dependence, 
there appear to be no functions $\xi$ that fulfill the associativity condition \eqref{3.4.6} and also reduce to \eqref{2.11}.  For instance, the choice 
\begin{equation}
\xi(g_2, g_1) =\frac{1}{2}m\Bigl( \Lambda_{b_1}\bs{a}_2 \cdot \Lambda_{b_1} \dot{R}_2\bs{a}_1 + \Lambda_{b_1} \bs{a}_2 \cdot \Lambda_{b_1} R_2 \dot{\bs{a}}_1  + \Lambda_{b_1} \dot{\bs{a}}_2 \cdot \Lambda_{b_1} R_2\bs{a}_1\Bigr)\label{3.4.9}
\end{equation}
satisfies  \eqref{3.4.6}, but it does not reduce \eqref{2.11}.  
On the other hand, the choices
\begin{equation}
\xi(g_2, g_1) = \frac{1}{2}m\Bigl(\Lambda_{b_1} \bs{a}_2 \cdot \Lambda_{b_1} R_2 \dot{\bs{a}}_1 - \Lambda_{b_1} \dot{\bs{a}}_2 \cdot \Lambda_{b_1} R_2\bs{a}_1\Bigr)\label{3.4.10}
\end{equation} 
and 
\begin{equation}
 \xi(g_2, g_1) =\frac{1}{2}m\Bigl( \Lambda_{b_1} \bs{a}_2 \cdot \Lambda_{b_1} \dot{R}_2\bs{a}_1 
 + \Lambda_{b_1} \bs{a}_2 \cdot \Lambda_{b_1} R_2 \dot{\bs{a}}_1  
      - \Lambda_{b_1} \dot{\bs{a}}_2 \cdot \Lambda_{b_1} R_2\bs{a}_1\Bigr)\label{3.4.11}
 \end{equation}
 do have the correct reduction to \eqref{2.11} but do not fulfill the associativity condition \eqref{3.4.6}. 
 
The results of this section show that in much the same way that inertial reference frames can be understood as the set of frames invariant under Galilei transformations, 
non-inertial frames can be understood as the set of frames that remains invariant under the Galilean line group. While both rotationally and linearly accelerating reference frames can 
be treated on an equal footing under the Galilean line group, rotational accelerations do appear to be quite different in character from linear accelerations in that it is 
in the absence of rotational accelerations that there exist extensions of the Galilean line group containing the central extensions of the Galilei group. 

We would expect the structure of rotational accelerations to be more complicated than that of linear accelerations because rotation matrices 
$R$, be they time dependent or not, do not commute whereas space translations $\bs{a}(t)$ always commute---therefore, so do 
boosts $\dot{\bs{a}}(t)$, accelerations $\ddot{\bs{a}}(t)$, etc. This suggests that we may face further difficulties if we were to follow the same approach to formulate 
a generalization of Lorentizian quantum mechanics to incorporate non-inertial reference frames because of the non-commutivity of the Lorentz boosts. On the other hand, 
since the Poincar\'e group has no central extensions, we would not expect an embedding problem analogous to what we have encountered here with the Galilei group. 

The above considerations lead us to conclude that either we must restrict our discussion to constant rotations or  abandon the embedding of central extensions 
of the Galilei group. The second choice would be justified if we were able to recover the representations of the centrally extended Galilei group in some limiting or approximate 
sense. This would mean that the usual Galilean quantum mechanics would come about as a limiting case of non-inertial quantum mechanics, rather like the way 
it does as a contraction of the Lorentzian quantum mechanics. Of the two choices, we will take the first in this paper. Therewith, we will henceforth consider only constant rotations.  

In the next section, we will take up the problem of constructing the unitary representations of the Galilean line group for constant rotations. Since the particular extension 
of the Galilean line group constructed above is non-central, our representations will not be projective. Rather, they will be higher cocycle representations in which, for example, 
associativity as well as group composition is satisfied only up to a phase factor.

\section{Representations of the Galilean line group}\label{sec4}
 In view of \eqref{2.9},  we must construct a cocycle representation of the Galilean 
line group. The simplest such choice is one in which the composition rule is given by
\begin{equation}
\hat{U}(g_2)\hat{U}(g_1)=e^{i\xi(g_2,g_1)}\hat{U}(g_2g_1)\label{4.0}
\end{equation}
where $\xi(g_2,g_1)=\frac{1}{2}m(\Lambda_{b_1}{\bs{a}}_2 \cdot R_2 \dot{\bs{a}}_1   - \Lambda_{b_1} \dot{\bs{a}}_2 \cdot R_2\bs{a}_1  )$ is the function \eqref{3.4.7} 
that defines the appropriate extension \eqref{3.4.4} of the Galilean line group.

We expect the $\hat{U}(g)$ to be unitary operators defined on a suitably constructed Hilbert space. However, since the Galilean line group is clearly non-compact, 
there are no solutions to the eigenvalue problem for the generators of this unitary representation. This is also the case for the unitary representations of the Galilei 
group; the vectors $\mid\bs{q}\rangle$ that we used to construct the representation \eqref{2.20} are not normalizable and so must be defined as elements of a vector 
space larger than the Hilbert space that carries the unitary representation.  The natural setting for dealing with the eigenvalue problem for the generators of unitary representations 
of non-compact groups is provided by the rigged 
Hilbert space formulation of quantum mechanics \cite{roberts,bohm,antoine}. A rigged Hilbert space is a triplet of spaces
\begin{equation}
\Phi\subset{\cal H}\subset\Phi^\times\label{4.0.1}
\end{equation}
where $\cal{H}$ is a Hilbert space, $\Phi$ is a dense subspace of ${\cal H}$ that is also a Fr\'echet space and $\Phi^\times$ is the topological antidual of $\Phi$ 
(the space of continuous antilinear functional on $\Phi$), equipped with its usual weak-$^*$ topology.  The particular choice of space $\Phi$ depends on the 
requirements of the physical problem at hand. 

Given a unitary representation of a Lie group, the space $\Phi$ must be constructed so that it carries a differentiable representation of the group. A canonical 
procedure for constructing such a $\Phi$, which leads to a rigged Hilbert space is given in \cite{sw}. When constructed this way, the differentiable representation carried 
by $\Phi$ is simply the restriction of the unitary representation in the Hilbert space ${\cal H}$ to $\Phi$. Therefore, we commonly denote the operators furnishing 
the differentiable representation in $\Phi$ also by the same symbol $\hat{U}(g)$. 

Given a differentiable representation of a Lie group $G$ in $\Phi$, there exists an associated dual representation  $\hat{U}^\times:\ \ \Phi^\times \times G\to\Phi^\times$, also  
differentiable, defined by the duality formula 
\begin{equation}
\langle \hat{U}(g)\phi\mid F\rangle=\langle\phi\mid\hat{U}^\times(g^{-1})F\rangle,\quad \phi\in\Phi,\ F\in\Phi^\times,\ g\in G\label{4.18}.
\end{equation}
For non-compact groups, it is really for this type of dual representation that familiar classical results of representation theory holds. For instance, 
irreducible representations of Abelian groups are one dimensional only if constructed as a representation in a suitable dual space $\Phi^\times$. 

When the representation furnished by $\hat{U}(g)$ in $\Phi$ 
is a cocycle representation, its dual representation in $\Phi^\times$ is also a cocyle representation. For instance, for a projective representation of the 
form \eqref{4.0}, applying \eqref{4.18} twice gives
\begin{equation}
\hat{U}^\times(g_2)\hat{U}^\times(g_1)=e^{-i\xi(g_1^{-1},g_2^{-1})}\hat{U}^\times(g_2g_1)\label{4.18b}.
\end{equation}
Using the explicit function \eqref{3.4.7}, for the Galilean line group we obtain 
\begin{equation}
\xi(g_1^{-1},g_2^{-1})=-\Lambda_{-b_1-b_2}\xi(g_2,g_1)\label{4.18c}
\end{equation}
and therewith
\begin{equation}
\hat{U}^\times(g_2)\hat{U}^\times(g_1)=e^{i\Lambda_{-b_1-b_2}\xi(g_2,g_1)}\hat{U}^\times(g_2g_1)\label{4.1}.
\end{equation}
Note the appearance of the shift operators in the phase factor of the dual cocycle representation, making it different from the phase factor in \eqref{4.0}. In contrast, for the Galilei group, 
using \eqref{2.11} we obtain $\omega(g_1^{-1},g_2^{-1})=-\omega(g_2,g_1)$ and therewith $\hat{U}^\times(g_2)\hat{U}^\times(g_1)=e^{i\omega(g_2,g_1)}\hat{U}^\times(g_2g_1)$, making the composition 
rule the same for a unitary projective representation and its dual representation. 
The difference between \eqref{4.0} and its dual representation \eqref{4.1} is due the non-central character of the extension of the Galilean line group given by \eqref{3.4.4}. Because of this difference, we will explicitly maintain 
the superscript $^\times$ when dealing with the dual representation. Strictly speaking, a distinction such as $\hat{H}$ and $\hat{H}^\times$ 
is also necessary between the generators of $\hat{U}(g)$ and $\hat{U}^\times(g)$. However, for a self-adjoint operator $\hat{A}$, the dual operator $\hat{A}^\times$ is an extension 
of $\hat{A}$ from $\Phi$ (or ${\cal H}$) to $\Phi^\times$ 
and for this reason, in the interest of notational economy, we will denote generators of the dual representation without the superscript $^\times$.

We will construct a cocycle representation of the  Galilean line group  of along the lines of 
Wigner's method of induced representations. Much the same way as for the Galilei group, this representation will be defined by the way operators $\hat{U}^\times(g)$ act on 
generalized velocity eigenvectors $\mid\bs{q}\rangle$, elements of $\Phi^\times$ of a rigged Hilbert space. 
The unitary Hilbert space representation can be obtained from this representation by way of the duality formula \eqref{4.18}. 
\emph{Although we will make use of the composition rule \eqref{4.1} to construct the representation,  as will be seen below in Section \ref{sec4.2}, 
the cocycle structure of the resulting representation is in fact more intricate than \eqref{4.1}}. 

\subsection{Construction of the representation}\label{sec4.1}
As is the case with projective representations of the Galilei 
group, we will induce representations of the Galilean line group from a representation of the Abelian subgroup $\{ (I, \bs{a}^0, b) \}$ of 
\emph{constant} spatial translations $\bs{a}^0$ and time translations $b$. For the sake of simplicity, we will restrict ourselves to spin-zero representations. Recall also that 
we have restricted ourselves to constant rotations. 

As  in \eqref{2.20} for the Galilei group, we will label the shared eigenvectors of this subgroup by $\mid \bs{q}_0 \rangle$. They are defined to transform under the 
subgroup $\hat{U}^\times(I,\bs{a}^0,b)$ exactly the same way as in \eqref{2.20}: 
\begin{equation}
\hat{U}^\times(I,\bs{a}^0,b)\mid \bs{q}_0\rangle=e^{im\bs{q}_0\cdot\bs{a}^0-iEb}\mid\bs{q}_0\rangle\label{4.2b}.
\end{equation}
The $\mid\bs{q}_0\rangle$ are also generalized eigenvectors of the generators of the operator subgroup $\hat{U}^\times(I,\bs{a}^0,b)$: 
\begin{eqnarray}
\hat{\bs{P}}\mid\bs{q}_0\rangle&\equiv&m\hat{\bs{Q}}\mid\bs{q}_0\rangle=m\bs{q}_0\mid\bs{q}_0\rangle\\
\hat{H}\mid\bs{q}_0\rangle&=&\frac{1}{2\hat{M}}\hat{\bs{P}}^2\mid\bs{q}_0\rangle=\frac{1}{2}m\hat{\bs{Q}}^2\mid\bs{q}_0\rangle=E\mid\bs{q_0}\rangle\label{4.3},
\end{eqnarray}
where $m$ and the generalized eigenvalues $\bs{q}_0\in\mathbb{R}^3$ have interpretation as the inertial mass and (constant) velocity of the particle as measured in an inertial reference frame. 

In order to determine the action of an arbitrary element $(R,\bs{a},b)$ of the line group on $\mid\bs{q}_0\rangle$, consider the 
following operator identities which readily follow from the composition rules for the group
 \eqref{3.1.5} and for the representation \eqref{4.1}:
 \begin{eqnarray}
 \hat{U}^\times(I,\bs{a}^0,b)\hat{U}^\times(R,(\bs{a}-\bs{a}^0),0)&=&e^{\frac{i}{2}m\Lambda_{-b}(\bs{a}^0\cdot\dot{\bs{a}})}\hat{U}^\times(R,\bs{a},b)\label{4.4a}\\
\hat{U}^\times(R,\Lambda_{-b}(\bs{a}-\bs{a}^0),0)\hat{U}^\times(I,R^{-1}\bs{a}^0,b)&=&e^{-\frac{i}{2}m\Lambda_{-b}(\dot{\bs{a}}\cdot\bs{a}^0)}\hat{U}^\times(R,\bs{a},b),
\label{4.4}
\end{eqnarray} 
where $\bs{a}^0$ is the zeroth order Taylor coefficient of the analytic function $\bs{a}$, i.e., $\bs{a}^0=\bs{a}(0)$. Combining \eqref{4.4a} and  \eqref{4.4}, we have 
\begin{equation}
\hat{U}^\times(I,\bs{a}^0,b)\hat{U}^\times(R,(\bs{a}-\bs{a}^0),0)=e^{im\Lambda_{-b}(\bs{a}^0\cdot\dot{\bs{a}})}\hat{U}^\times(R,\Lambda_{-b}(\bs{a}-\bs{a}^0),0)\hat{U}^\times(I,R^{-1}\bs{a}^0,b)\label{4.5}.
\end{equation}
Evaluating \eqref{4.5} at $b=0$ and using \eqref{4.2b}, we obtain
\begin{eqnarray}
\hat{U}^\times(I,\bs{a}^0,0)\hat{U}^\times(R,(\bs{a}-\bs{a}^0),0)\mid\bs{q}_0\rangle&=&e^{im\bs{a}^0\cdot{\dot{\bs{a}}}}e^{imR^{-1}\bs{a}^0\cdot\bs{q}_0}\,\hat{U}^\times
(R,(\bs{a}-\bs{a}_0),0)\mid\bs{q}_0\rangle\nonumber\\
&=&e^{im\bs{a}^0\cdot(R\bs{q}_0+\dot{\bs{a}})}\hat{U}^\times(R,(\bs{a}-\bs{a}^0),0)\mid\bs{q}_0\rangle.\label{4.6}
\end{eqnarray}
This shows that $\hat{U}^\times(R,(\bs{a}-\bs{a}^0),0)\mid\bs{q}_0\rangle$ is an eigenvector of the constant spacial translation subgroup $\hat{U}^\times(I,\bs{a}^0,0)$ with eigenvalue 
$e^{im\bs{a}^0\cdot(R\bs{q}_0+\dot{\bs{a}})}$. 
This property suggests that we \emph{define} generalized eigenvectors associated with \emph{time dependent velocities} $\bs{q}$ by
\begin{equation}
\mid\bs{q}\rangle:=\frac{1}{N(\bs{q}_0,\bs{q})}e^{-i\zeta(\bs{q},\bs{a})}\,\hat{U}^\times(R,\bs{a},0)\mid\bs{q}_0\rangle\label{4.7},
\end{equation}
where $\bs{q}(t)=R\bs{q}_0+\dot{\bs{a}}(t)$ and $N(\bs{q}_0,\bs{q})$ is a number to be determined by a normalization convention for vectors $\mid\bs{q}\rangle$.
(Incidentally, the transformation $\bs{q}_0\to\bs{q}(t)=R\bs{q}_0+\dot{\bs{a}}(t)$ is another reason to restrict ourselves to constant rotations, as this will not furnish a homomorphism in the case $\dot{R} \neq 0$.)
 The 
$\zeta(\bs{q},\bs{a})$ that appears in the phase factor must be determined so that \eqref{4.7} reduces to \eqref{4.2b} for $\bs{a}=\bs{a}^0$ and to the Galilei group representation \eqref{2.20} when 
$\dot{\bs{a}}=\bs{v}$, a constant. These criteria suggest that $\zeta(\bs{q},\bs{a})$ be of the form 
\begin{equation}
\zeta(\bs{q},\bs{a})=m(\bs{q}\cdot\bs{a}-\frac{1}{2}\bs{a}\cdot\dot{\bs{a}})\label{4.8}.
\end{equation} 
Thus, we write \eqref{4.7} as 
\begin{equation}
\hat{U}^\times(R,\bs{a},0)\mid\bs{q}_0\rangle=N(\bs{q}_0,\bs{q})e^{im(\bs{q}\cdot\bs{a}-\frac{1}{2}\bs{a}\cdot{\dot{\bs{a}}})}\mid\bs{q}\rangle\label{4.9}.
\end{equation}
In particular, if we set $\bs{q}_0=0$ in \eqref{4.9}, then it follows
\begin{equation}
\hat{U}^\times(R,\bs{a},0)\mid\bs{0}\rangle=N(\bs{q})e^{\frac{i}{2}m\bs{q}\cdot\bs{a}}\mid\bs{q}\rangle\label{4.9b},
\end{equation}
where $\bs{q}=R\bs{q}_0+\dot{\bs{a}}=\dot{\bs{a}}$. This means that for any time dependent velocity, there exists a spatial 
translation function $\bs{a}_{\bs{q}}$ where $\dot{\bs{a}_{\bs{q}}}=\bs{q}$ and therewith a 
line group element
\begin{equation}
g_{\bs{q}}:=(I,\bs{a}_{\bs{q}},0),\quad \dot{\bs{a}_{\bs{q}}}=\bs{q}\label{4.9bb},
\end{equation}
such that the generalized eigenvector $\mid\bs{q}\rangle$ can be obtained by applying the boost operator $\hat{U}^\times(g_{\bs{q}})$ on the rest vector $\mid\bs{0}\rangle$. It is seen from \eqref{4.9b} that for a given $\mid\bs{q}\rangle$, all operators $\hat{U}^\times(R,\bs{a}_{\bs{q}},0)=\hat{U}^\times(g_{\bs{q}})\hat{U}^\times(R,\bs{0},0)$ yield the same result--our little group is still $SO(3)$.  Moreover, 
the condition $\dot{\bs{a}_{\bs{q}}}=\bs{q}$ defines $\bs{a}_{\bs{q}}$ only up to a constant which can be set to zero by choosing $\bs{a}_{\bs{q}}(0)=0$. 
With these conventions, we may define $\mid\bs{q}\rangle$ with the use of the rotation free operator $\hat{U}^\times(g_{\bs{q}})$: 
\begin{equation}
\mid{\bs{q}}\rangle:=\frac{1}{N(\bs{q})}e^{-\frac{i}{2}m\bs{q}\cdot\bs{a}_{\bs{q}}}\,\hat{U}^\times(g_{\bs{q}})\mid\bs{0}\rangle\label{4.9c}.
\end{equation}
The action of operators $\hat{U}^\times(R,\bs{a},0)$ on arbitrary velocity vectors $\mid\bs{q}\rangle$ can be determined from \eqref{4.1}, \eqref{4.9b} and \eqref{4.9c}: 
\begin{eqnarray}
\hat{U}^\times(R,\bs{a},0)\mid\bs{q}\rangle&=&\frac{1}{N(\bs{q})}e^{-\frac{i}{2}m\bs{q}\cdot{\bs{a}_{\bs{q}}}}\hat{U}^\times(R,\bs{a},0)\hat{U}^\times(g_{\bs{q}})\mid\bs{0}\rangle\nonumber\\
&=&\frac{1}{N(\bs{q})}e^{-\frac{i}{2}m\bs{q}\cdot{\bs{a}_{\bs{q}}}}e^{\frac{i}{2}m(\bs{a}\cdot R\bs{q}-\dot{\bs{a}}\cdot R\bs{a}_{\bs{q}})}\hat{U}^\times(R,\bs{a}+R\bs{a}_{\bs{q}},0)\mid\bs{0}\rangle\nonumber\\
&=&\frac{N(\bs{q}')}{N(\bs{q})}e^{\frac{i}{2}m(\bs{q}'\cdot(\bs{a}+R\bs{a}_{\bs{q}})-\bs{q}\cdot{\bs{a}_{\bs{q}}}+(\bs{a}\cdot R\bs{q}-\dot{\bs{a}}\cdot R\bs{a}_{\bs{q}}))}\mid\bs{q}'\rangle\nonumber\\
&=&\frac{N(\bs{q}')}{N(\bs{q})}e^{im(\bs{q}'\cdot\bs{a}-\frac{1}{2}\bs{a}\cdot\dot{\bs{a}})}\mid\bs{q}'\rangle,\label{4.10}
\end{eqnarray}
where 
\begin{equation}
\bs{q}'=\bs{0}+\dot{(\bs{a}+R\bs{a}_{\bs{q}})}=R\bs{q}+\dot{\bs{a}}\label{4.10b}.
\end{equation}

We now turn to the time translations subgroup. Setting $\bs{a}^0=0$ and $\bs{a}=\bs{a}_{\bs{q}}$ in \eqref{4.5}, applying the resulting operator equality on the rest vector $\mid\bs{0}\rangle$,   
and using \eqref{4.9c}, we obtain
\begin{eqnarray}
\hat{U}^\times(I,\bs{0},b)\hat{U}^\times(R,\bs{a}_{\bs{q}},0)\mid\bs{0}\rangle&=&\hat{U}^\times(R,\Lambda_{-b}\bs{a}_{\bs{q}},0)\hat{U}^\times(I,\bs{0},b)\mid\bs{0}\rangle\nonumber\\
\hat{U}^\times(I,\bs{0},b)\mid\bs{q}\rangle&=&\frac{1}{N(\bs{q})}e^{-\frac{i}{2}m\bs{q}
\cdot\bs{a}_{\bs{q}}}\hat{U}^\times(R,\Lambda_{-b}\bs{a}_{\bs{q}},0)\hat{U}^\times(I,\bs{0},b)\mid\bs{0}\rangle.
\label{4.11}
\end{eqnarray}
Since the action of operators $\hat{U}^\times(R,\bs{a}_{\bs{q}},0)$ is known from \eqref{4.10}, this equality tells us that the 
action of time translation operators $\hat{U}^\times(I,\bs{0},b)$ on velocity vectors $\mid\bs{q}\rangle$ is determined by the action 
of $\hat{U}^\times(I,\bs{0},b)$ on the rest vector $\mid\bs{0}\rangle$. We appeal to \eqref{4.2b} to infer
\begin{equation}
\hat{U}^\times(I,\bs{0},b)\mid\bs{0}\rangle=e^{-iwb}\mid\bs{0}\rangle\label{4.11b},
\end{equation}
where $w$ is the value of the parameter $E$ in the rest frame, i.e., it is the internal energy of the system. Therewith, \eqref{4.11} becomes 
\begin{equation}
\hat{U}^\times(I,\bs{0},b)\mid\bs{q}\rangle=\frac{N(\Lambda_{-b}\bs{q})}{N(\bs{q})}e^{\frac{i}{2}m(\Lambda_{-b}-1)\bs{q}
\cdot\bs{a}_{\bs{q}}}e^{-iwb}\mid\Lambda_{-b}\bs{q}\rangle\label{4.12}.
\end{equation}

The general transformation of an arbitrary velocity state under Galilean line group readily follows from \eqref{4.10} and \eqref{4.12}: 
\begin{eqnarray}
\hat{U}^\times(R,\bs{a},b)\mid\bs{q}\rangle&=&\hat{U}^\times(I,\bs{0},b)\hat{U}^\times(R,\bs{a},0)\mid\bs{q}\rangle\nonumber\\
&=&\frac{N(\Lambda_{-b}\bs{q}')}{N(\bs{q})}e^{im(\bs{q}'\cdot\bs{a}-\frac{1}{2}\bs{a}\cdot{\dot{\bs{a}}}+\frac{1}{2}(\Lambda_{-b}-1)\bs{q}'\cdot{\bs{a}}_{\bs{q}'})}e^{-iwb}\mid\Lambda_{-b}\bs{q}'\rangle.
\label{4.14}
\end{eqnarray}
where, as before, $\bs{q}'=R\bs{q}+\dot{\bs{a}}$ and $\bs{a}_{\bs{q}'}$ is the standard boost that maps  the zero vector to $\bs{q}'$ ($\dot{\bs{a}_{\bs{q}'}}=\bs{q}'$).

It remains to determine the normalization constant $N(\bs{q})$. As stated above, the vectors $\mid\bs{q}\rangle$ that we have used to 
construct the representation are not square integrable and as such they are not defined  as elements of a Hilbert space. Instead, they should be defined 
as elements of the dual space $\Phi^\times$ of a suitably constructed rigged Hilbert space of the form \eqref{4.0.1}. 
 Under these conditions, all $\mid\psi\rangle\in\Phi$ have a basis vector expansion with regard to $\mid\bs{q}\rangle$:
\begin{equation}
\psi=\int d\mu(\bs{q})\mid\bs{q}\rangle\langle\bs{q}\mid\psi\rangle\label{4.15},
\end{equation}
where 
\begin{equation}
\psi(\bs{q})=\langle\bs{q}\mid\psi\rangle\label{4.15b}
\end{equation}
is the velocity wave function. 
If we take the basis vectors to be Dirac  delta normalized,
\begin{equation}
\langle\bs{q'}\mid\bs{q}\rangle=\rho(\bs{q})\delta(\bs{q}-\bs{q}')\label{4.16},
\end{equation}
for some $\mu$-measurable function $\rho$ such that $\int d\mu(\bs{q}') \rho(\bs{q}')\delta(\bs{q}-\bs{q}')f(\bs{q}')=f(\bs{q})$, then $\psi$ are square integrable:
\begin{equation}
\langle\psi\mid\psi\rangle=\int d\mu(\bs{q})\psi^*(\bs{q})\psi(\bs{q})\label{4.17}.
\end{equation}
Since the $\bs{q}$ are functions of time, it should be noted that \eqref{4.15}-\eqref{4.17} must be understood as defined for each instance of time. For instance, 
$\delta(\bs{q}-\bs{q}')(t)=\delta(\bs{q}(t)-\bs{q}'(t))$ and $\langle\psi\mid\psi\rangle=\int d\mu(\bs{q}(t))\psi^*(\bs{q}(t))\psi(\bs{q}(t))$.

A representation of the Galilean line group on the space $\Phi$ of \eqref{4.0.1} can be obtained from the transformation formula \eqref{4.14}, the definition of the 
wave function \eqref{4.15b}  and the duality formula \eqref{4.18}:
\begin{eqnarray}
\Bigl(\hat{U}(R,\bs{a},b)\psi\Bigr)(\bs{q})&=&\langle\psi\mid \hat{U}^\times(\Lambda_{-b}R^{-1},-\Lambda_{-b}(R^{-1}\bs{a}),-b)\bs{q}\rangle^*\nonumber\\
&=&\frac{N^*(\Lambda_b\tilde{\bs{q}})}{N^*(\bs{q})}e^{\frac{i}{2}m\{(\Lambda_b+1)\bs{q}\cdot\Lambda_{-b}\bs{a}-
\bs{a}\cdot{\dot{\bs{a}}}-(\Lambda_b-1)\bs{q}\cdot\bs{a}_{\bs{q}}-(\Lambda_{-b}\dot{\bs{a}}\cdot\bs{a}_{\bs{q}}-\dot{\bs{a}}\cdot\Lambda_b\bs{a}_{\bs{q}})\}}e^{-iwb}
\psi(\Lambda_b\tilde{\bs{q}}),\nonumber\\
\label{4.19}
\end{eqnarray}
where $\tilde{\bs{q}}=R^{-1}\bs{q}-R^{-1}\Lambda_{-b}\dot{\bs{a}}$.

Substituting \eqref{4.19} in \eqref{4.17} gives us
\begin{equation}
\langle \hat{U}(g)\psi\mid \hat{U}(g)\psi\rangle=\int d\mu(\bs{q})\left|\frac{N(\Lambda_b\tilde{\bs{q}})}{N(\bs{q})}\right|^2\psi^*(\Lambda_b\tilde{\bs{q}})\psi(\Lambda_b{\tilde{\bs{q}}})\label{4.20}.
\end{equation}
Now, should it be possible to choose the integration measure and normalization constant such that 
\begin{equation}
d\mu(\bs{q})\left|\frac{N(\Lambda_b\tilde{\bs{q}})}{N(\bs{q})}\right|^2=d\mu(\Lambda_b\tilde{\bs{q}}),\label{4.21}
\end{equation}
then the integrals of \eqref{4.17} and \eqref{4.20} will yield the same value. Since $\Phi$ is dense in the Hilbert space, this means that the representation will extend to a
unitary representation in the Hilbert space $\cal H$.  Noting the invariance of  the usual Lebesgue measure on $\mathbb{R}^3$ under the Galilean line group, 
it is simplest to let $d\mu(\bs{q})=d^3\bs{q}$ and therewith $N(\bs{q})=1$ and $\rho(\bs{q})=1$. With these choices, the construction of the unitary representation of the  Galilean line group is complete. The representation is defined either by the transformation formula for the generalized eigenvectors of momentum operator, 
\begin{equation}
\hat{U}^\times(R,\bs{a},b)\mid\bs{q}\rangle=e^{i\xi(g;\bs{q})}\mid\Lambda_{-b}\bs{q}'\rangle\label{4.22},
\end{equation}
where 
\begin{eqnarray}
\xi(g;\bs{q})&=&m(\bs{q}'\cdot\bs{a}-\frac{1}{2}\bs{a}\cdot{\dot{\bs{a}}}+\frac{1}{2}(\Lambda_{-b}-1)\bs{q}'\cdot{\bs{a}}_{\bs{q}'})-wb\nonumber\\
\bs{q}'&=&R\bs{q}+\dot{\bs{a}},\label{4.22b}
\end{eqnarray}
or by the transformation formula for the $L^2$-functions, 
\begin{equation}
\Bigl(\hat{U}(R,\bs{a},b)\psi\Bigr)(\bs{q})=e^{-i\xi(g^{-1};\bs{q})}\psi(\Lambda_{b}\tilde{\bs{q}})\label{4.23},
\end{equation}
where
\begin{eqnarray}
\xi(g^{-1};\bs{q})&=&-\frac{1}{2}m\left\{(\Lambda_b+1)\bs{q}\cdot\Lambda_{-b}\bs{a}-
\bs{a}\cdot{\dot{\bs{a}}}-(\Lambda_b-1)\bs{q}\cdot\bs{a}_{\bs{q}}-(\Lambda_{-b}\dot{\bs{a}}\cdot\bs{a}_{\bs{q}}-\dot{\bs{a}}\cdot\Lambda_b\bs{a}_{\bs{q}})\right\}+wb\nonumber\\
\tilde{\bs{q}}&=&(\Lambda_{-b}R^{-1})\bs{q}-\Lambda_{-b}\dot{(R^{-1}\bs{a})}=R^{-1}\bs{q}-R^{-1}\Lambda_{-b}\dot{\bs{a}}.\label{4.23b}
\end{eqnarray}

This unitary representation of the Galilean line group contains the unitary projective representation of the Galilei group \eqref{2.20} as a subrepresentation. To see this, we follow the procedure 
used to obtain \eqref{3.4.8} from \eqref{3.4.7}: let $\bs{a}(t)=\bs{a}^0+\bs{v}t$, $\dot{\bs{a}}=\bs{v}$ and evaluate the resulting expression at $t=0$, taking care to note $\Lambda_b\bs{a}(0)=\bs{a}(b)$. 
Thus,  letting $\bs{q}=\bs{q}_0$, $\bs{a}_{\bs{q}_0}=\bs{q}_0t$, and $\bs{a}(t)=\bs{a}_0+\bs{v}t$ in \eqref{4.22},
\begin{eqnarray}
\hat{U}^\times(R, \bs{a}(t), b)\mid \bs{q}_0\rangle &= &
e^{im((R\bs{q}_0+\bs{v})\cdot(\bs{a}^0+\bs{v}t)-\frac{1}{2}(\bs{a}^0+\bs{v}t)\cdot{{\bs{v}}}+\frac{1}{2}(\Lambda_{-b}-1)(R\bs{q}+\bs{v}\cdot(R\bs{q}+\bs{v})t))}
e^{-iwb}\mid R\bs{q}_0+\bs{v}\rangle\nonumber\\
&=&e^{im\left((R\bs{q}_0+\bs{v})\cdot\bs{a}^0-\frac{1}{2}\bs{a}^0\cdot\bs{v}+((R\bs{q}_0+\bs{v})\cdot\bs{v}-\frac{1}{2}\bs{v}^2)t\right)}
e^{-i(w+\frac{1}{2}m(R\bs{q}_0+\bs{v})^2)b}\mid R\bs{q}_0+\bs{v}\rangle
\end{eqnarray}
and thus 
\begin{eqnarray}
\left. \hat{U}^\times(R, \bs{a}(t), b)\mid \bs{q}_0 \rangle \right|_{t=0} &=& e^{im( \bs{a}^0 \cdot \bs{q}_0'- \frac{1}{2}\bs{a}^0\cdot\bs{v})}e^{-iE'b} \mid \bs{q}'_0\rangle\label{4.24},
\end{eqnarray}
where $\bs{q}_0'=R\bs{q}_0+\bs{v}$ and $E'=w+\frac{1}{2}m{\bs{q}'_0}^2$. Hence, \eqref{4.24} is exactly the Galilei group representation. This embedding of the Galilei group representation leads us to the interpretation that the parameter $m$ in the line group representation \eqref{4.22} or \eqref{4.23} 
is the \emph{inertial mass}.

\subsection{The cocycle structure of the representation}\label{sec4.2}
Recall that our construction of the unitary representation of the Galilean line group made use of the composition rule \eqref{4.1}. However, since the group extension 
corresponding to the phase factor of \eqref{4.1} is not central, we expect physically relevant representations of the Galilean line group to have a cocycle structure that is more complicated than that of a projective representation. Discussions of cochains, coboundaries and cocycles,  the central ideas involving cocycle representations, can be found in many books on algebraic topology or mathematical physics. For instance, 
see \cite{hatcher}.

In view of \eqref{4.24}, we expect the group operators of \eqref{4.22} or \eqref{4.23} to furnish a homomorphism only up to a phase factor. In fact, composing the two operators $\hat{U}^\times(R_2,\bs{a}_2,b_2)$ and $\hat{U}^\times(R_1,\bs{a}_1,b_1)$ using \eqref{4.22}, we 
obtain 
\begin{equation}
\hat{U}^\times(g_2)\hat{U}^\times(g_1)\mid\bs{q}\rangle=e^{i\xi(g_1;\bs{q})}e^{i\xi(g_2;\Lambda_{-b_1}\bs{q}')}\mid\Lambda_{-b_2-b_1}\bs{q}''\rangle,\label{4.1.1}
\end{equation}
where
\begin{equation}
{\bs{q}}''=R_2R_1\bs{q}+R_2\dot{\bs{a}}_1+\Lambda_{b_1}\dot{\bs{a}}_2.\label{4.1.2}
\end{equation}
On the other hand, using \eqref{4.22} for $\hat{U}^\times(g_2g_1)$, 
\begin{equation}
\hat{U}^\times(g_2g_1)\mid\bs{q}\rangle=e^{i\xi(g_2g_1;\bs{q})}\mid\Lambda_{-b_1-b_2}\bs{q}''\rangle.\label{4.1.3}
\end{equation}
Therefore, 
\begin{equation}
\hat{U}^\times(g_2)\hat{U}^\times(g_1)\mid\bs{q}\rangle=e^{i\xi(g_2,g_1;\bs{q})}\hat{U}^\times(g_2g_1)\mid\bs{q}\rangle\label{4.1.4},
\end{equation}
where 
\begin{equation}
\xi(g_2,g_1;\bs{q}):=\xi(g_1;\bs{q})+\xi(g_2;\Lambda_{-b_1}\bs{q}')-\xi(g_2g_1;\bs{q})\label{4.1.4b}.
\end{equation}
Substituting from \eqref{4.22b} and after some algebra, \eqref{4.1.4b} becomes
\begin{equation}
\xi(g_2,g_1;\bs{q})=\xi(g_2,g_1)+\frac{1}{2}m(\Lambda_{-b_1}-1)(\Lambda_{b_1}\bs{a}_2\cdot\dot{\bs{a}}_{\bs{q}''}-\Lambda_{b_1}\dot{\bs{a}}_2\cdot{\bs{a}_{\bs{q}''}})\label{4.1.5}.
\end{equation}
Note that the second term of \eqref{4.1.5} also has the same structure as the function $\xi(g_2,g_1)$. In fact, writing out $\bs{a}_{\bs{q}''}$ in terms of $\bs{a}_{\bs{q}}$ with the use of \eqref{4.1.2}, 
\begin{equation}
\bs{a}_{\bs{q}''}=R_2R_1\bs{a}_{\bs{q}}+R_2\bs{a}_1+\Lambda_{b_1}\bs{a}_2\label{4.1.6}.
\end{equation}
Thus,
\begin{eqnarray}
\Lambda_{b_1}\bs{a}_2\cdot\dot{\bs{a}}_{\bs{q}''}-\Lambda_{b_1}\dot{\bs{a}}_2\cdot{\bs{a}_{\bs{q}''}}&=&\Lambda_{b_1}\bs{a}_2\cdot R_2(\dot{\bs{a}}_1+R_1\dot{\bs{a}}_{\bs{q}})-\Lambda_{b_1}\dot{\bs{a}}_2\cdot R_2(\bs{a}_1+R_1\bs{a}_{\bs{q}})\nonumber\\
&=&\frac{2}{m}\xi(g_2,g_1g_{\bs{q}}).\label{4.1.7}
\end{eqnarray}
and therewith \eqref{4.1.5} reads 
\begin{equation}
\xi(g_2,g_1;\bs{q})=\xi(g_2,g_1)+(\Lambda_{-b_1}-1)\xi(g_2,g_1g_{\bs{q}})\label{4.1.8}.
\end{equation}
For the homomorphism property to hold for the representation, this function must either vanish or be of the form 
\begin{equation}
\xi(g_2,g_1;\bs{q})=\phi(g_2g_1;\Lambda_{-b_2-b_1}\bs{q}'')-\phi(g_1;\bs{q})-\phi(g_2;\Lambda_{-b_1}\bs{q}')\label{4.1.8b}
\end{equation}
for some function $\phi$ (i.e., a coboundary) so that the phase factor of \eqref{4.1.4} can be removed by a redefinition of the operators $\hat{U}^\times(g)$. That neither condition is satisfied 
readily follows from the fact that the representation contains a projective representation of the Galilei subgroup.  While the Galilean subrepresentation is projective, 
the appearance of the velocity variable $\bs{q}$ in \eqref{4.1.8} shows that the Galilean line group representation we have constructed is not a projective representation but a more general cocycle 
representation. 

Whether our representation is  a two cocycle representation can be determined by checking if associativity  holds for the $\hat{U}^\times(g)$.  To that end,  using \eqref{4.1.4}, for $\hat{U}^\times(g_3)\Bigl(\hat{U}^\times(g_2)\hat{U}^\times(g_1)\Bigr)$ we obtain
\begin{eqnarray}
\hat{U}^\times(g_3)\Bigl(\hat{U}^\times(g_2)\hat{U}^\times(g_1)\Bigr)\mid\bs{q}\rangle&=&e^{i\xi(g_2,g_1;\bs{q})}\hat{U}^\times(g_3)\hat{U}^\times(g_2g_1)\mid\bs{q}\rangle\nonumber\\
&=&e^{i\xi(g_2,g_1;\bs{q})}e^{i\xi(g_3,g_2g_1;\bs{q})}\hat{U}^\times(g_3g_2g_1)\mid\bs{q}\rangle,
\label{4.1.9}
\end{eqnarray}
while for $\Bigl(\hat{U}^\times(g_3)\hat{U}^\times(g_2)\Bigr)\hat{U}^\times(g_1)$, 
\begin{eqnarray}
\Bigl(\hat{U}^\times(g_3)\hat{U}^\times(g_2)\Bigr)\hat{U}^\times(g_1)\mid\bs{q}\rangle&=&e^{i\xi(g_1;\bs{q})}\Bigl(\hat{U}^\times(g_3)\hat{U}^\times(g_2)\Bigr)\mid\Lambda_{-b_1}\bs{q}'\rangle\nonumber\\
&=&e^{i\xi(g_1;\bs{q})}e^{i\xi(g_3,g_2;\Lambda_{-b_1}\bs{q}')}\hat{U}^\times(g_3g_2)\mid\Lambda_{-b_1}\bs{q}'\rangle\nonumber\\
&=&e^{i\xi(g_3,g_2;\Lambda_{-b_1}\bs{q}')}\hat{U}^\times(g_3g_2)\hat{U}^\times(g_1)\mid\bs{q}\rangle\nonumber\\
&=&e^{i\xi(g_3,g_2;\Lambda_{-b_1}\bs{q}')}e^{i\xi(g_3g_2,g_1;\bs{q})}\hat{U}^\times(g_3g_2g_1)\mid\bs{q}\rangle,
\label{4.1.10}
\end{eqnarray}
where we have used \eqref{4.22} to obtain the first and third equalities. 
Hence, 
\begin{equation}
\Bigl(\hat{U}^\times(g_3)\hat{U}^\times(g_2)\Bigr)\hat{U}^\times(g_1)\mid\bs{q}\rangle=e^{i\xi(g_3,g_2,g_1;\bs{q})}\hat{U}^\times(g_3)\Bigl(\hat{U}^\times(g_2)\hat{U}^\times(g_1)\Bigr)\mid\bs{q}\rangle\label{4.1.11},
\end{equation}
where
\begin{equation}
\xi(g_3,g_2,g_1;\bs{q})=\xi(g_3,g_2;\Lambda_{-b_1}\bs{q}')+\xi(g_3g_2,g_1;\bs{q})-\xi(g_2,g_1;\bs{q})-\xi(g_3,g_2g_1;\bs{q})\label{4.1.12}.
\end{equation}
For the associativity of the representation to hold,  this function must be a coboundary (i.e., vanish or the phase factor of \eqref{4.1.11} must be able to be removed by a redefinition  of 
operators $\hat{U}^\times(g)$). Substituting \eqref{4.1.8} for each term on the right hand side of \eqref{4.1.12} and using \eqref{3.4.6}, it can be shown that the function
$\xi(g_3,g_2,g_1;\bs{q})$ does not vanish and that the phase factor of \eqref{4.1.11} cannot be removed by redefining the group operators. 

These considerations show that the Galilean line group representation we have obtained is at least a three cocycle representation. Three cocycle representations are not common in physics, but they have been noted to appear in a few examples \cite{jackiw, grossman, nesterov, hou}. The complicated cocycle structure of our representation suggests the Galilean line group has a rich cohomology, an investigation that we intend to undertake in a future study. It should also be mentioned that there are other representations of the Galilean line group with different cocycle properties. For instance, the representation defined by 
\begin{equation}
\hat{U}^\times(R, \bs{a}, b)  \mid \bs{q} \rangle 
= e^{im\left( \Lambda_{-b}( \bs{q}' \cdot\bs{a} ) - \frac{1}{2}  \Lambda_{-b}(\bs{a} \cdot \dot{\bs{a}} ) \right)}e^{-iEb}  \mid \Lambda_{-b} \bs{q}' \rangle = \Lambda_{-b} \left( e^{-iEb}
e^{im\left(( \bs{q}' \cdot\bs{a} ) - \frac{1}{2} (\bs{a} \cdot \dot{\bs{a}} ) \right)}\mid \bs{q}' \rangle \right)
\end{equation}
where $E$ is a constant is a projective representation fulfilling exactly \eqref{4.1}. However, this representation does not contain a projective representation of the Galilei group. 

In passing, we remark that our initial anticipation of a projective representation satisfying \eqref{4.1}, wrong in light of \eqref{4.1.8}, does not lead to inconsistent results because in all of the instances where we made use of \eqref{4.1}, such as \eqref{4.4}, the correct phase factor \eqref{4.1.8} coincides with that of \eqref{4.1}. 

\subsection{The Hamiltonian}\label{sec4.3}
As is the case in quantum mechanics in inertial reference frames, we define the Hamiltonian as the generator of the subgroup $\hat{U}(I,\bs{0},b)$ of time translations: 
\begin{equation}
\hat{H}:=i\left.\frac{d\hat{U}(I,\bs{0},b)}{db}\right|_{b=0}\label{4.2.1}.
\end{equation}
Before we evaluate this derivative, first note that there is always the term $-iwb$ in the exponent of the transformation formula 
 \eqref{4.22} or \eqref{4.23}. For definiteness, let us work with \eqref{4.23} for the remainder of this section. 
 The appearance of this common factor  suggests the existence of an operator $\hat{V}$ that may be interpreted as a \emph{internal energy} (cf. last equality of \eqref{2.16}):
\begin{equation}
\left(\hat{V}\psi\right)(\bs{q})=w\psi(\bs{q})\label{4.2.2}.
\end{equation}
Indeed, from \eqref{4.11b} we note that $\hat{V}$ coincides with the Hamiltonian in the rest frame. It generates a one parameter group of unitary operators
\begin{equation}
\hat{U}_{int}(b)=e^{-i\hat{V}b}\label{4.2.3}.
\end{equation}
Both $\hat{U}_{int}(b)$ and its generator $\hat{V}$ commute with all other operators associated with the Galilean line group representation. 

The realization of the operators $\hat{U}(I,\bs{0},b)$ in the $L^2(\mathbb{R}^3)$ function space can be obtained from \eqref{4.23} by setting $R=I$ and 
$\bs{a}=0$:
\begin{eqnarray}
\Bigl(\hat{U}(I,\bs{0},b)\psi\Bigr)(\bs{q})&=&e^{-i\frac{1}{2}m(\Lambda_b-1)\bs{q}\cdot\bs{a}_{\bs{q}}}e^{-iwb}\psi(\Lambda_b\bs{q})\nonumber\\
&=&e^{i\frac{1}{2}m\bs{q}\cdot\bs{a}_{\bs{q}}}\hat{U}_{int}(b)\Lambda_b\Bigl( e^{-i\frac{1}{2}m\bs{q}\cdot\bs{a}_{\bs{q}}}\psi(\bs{q})\Bigr),\label{4.2.4}
\end{eqnarray}
where the second equality follows from the fact that the dependence of velocity wave function $\psi(\bs{q})$ on the parameter $t$ is entirely determined by 
the velocity variable $\bs{q}$, i.e., $\Lambda_b\psi(\bs{q})=\psi(\Lambda_b\bs{q})$. 

The appearance of the shift operator $\Lambda_b$ in \eqref{4.2.4} suggests that 
 the Hamiltonian must contain the derivative operator, the generator of the automorphism group $\{\Lambda_b\}$, along with $\hat{V}$. 
In fact, differentiating \eqref{4.2.4} with respect to $b$, 
\begin{equation}
\left(\frac{d\hat{U}(I,\bs{0},b)}{db}\psi\right)(\bs{q})=e^{i\frac{1}{2}m\bs{q}\cdot\bs{a}_{\bs{q}}}\left(\frac{d\hat{U}_{int}(b)}{db}\Lambda_b+\hat{U}_{int}(b)\frac{d\Lambda_b}{db}\right)\Bigl( e^{-i\frac{1}{2}m\bs{q}\cdot\bs{a}_{\bs{q}}}\psi(\bs{q})\Bigr)
\label{4.2.7}
\end{equation}
We can easily compute $\frac{d}{db} \Lambda_{b} $:
\begin{eqnarray}
\frac{d}{db}\Lambda_{b} f(t) &=& \lim_{\varepsilon \ra 0} \frac{f(t+b+\varepsilon) - f(t + b)}{\varepsilon}\nonumber\\
&=&\Lambda_{b} \lim_{\varepsilon \ra 0} \frac{f(t+\varepsilon) - f(t)}{\varepsilon}\nonumber\\
&=&\Lambda_{b}\frac{d}{dt}f(t).\label{4.2.8}
\end{eqnarray}
This says that differentiation operator is the generator of the one parameter automorphism group $\Lambda_b$. Hence, if we denote by $\hat{D}_t$ the differentiation operator with respect to time, then
\begin{equation}
\Lambda_b=e^{b\hat{D}_t}\label{4.2.9}.
\end{equation}
Substituting \eqref{4.2.3} and \eqref{4.2.8} in \eqref{4.2.7},
\begin{eqnarray}
i\left(\frac{d\hat{U}(I,\bs{0},b)}{db}\psi\right)(\bs{q})&=&e^{i\frac{1}{2}m\bs{q}\cdot\bs{a}_{\bs{q}}}(\hat{V}+i\hat{D}_t)\hat{U}_{int}(b)\Lambda_b\Bigl( e^{-i\frac{1}{2}m\bs{q}\cdot\bs{a}_{\bs{q}}}\psi(\bs{q})\Bigr)\nonumber\\
&=&e^{i\frac{1}{2}m\bs{q}\cdot\bs{a}_{\bs{q}}}\hat{U}_{int}(b)\Lambda_b(\hat{V}+i\hat{D}_t)\Bigl( e^{-i\frac{1}{2}m\bs{q}\cdot\bs{a}_{\bs{q}}}\psi(\bs{q})\Bigr).
\label{4.2.10}
\end{eqnarray}
Hence, 
\begin{eqnarray}
\left(\hat{H}\psi\right)(\bs{q})&:=&i\left(\left.\frac{d\hat{U}(I,\bs{0},b)}{db}\right|_{b=0}\psi\right)(\bs{q})\nonumber\\
&=&e^{i\frac{1}{2}m\bs{q}\cdot\bs{a}_{\bs{q}}}(\hat{V}+i\hat{D}_t)\Bigl( e^{-i\frac{1}{2}m\bs{q}\cdot\bs{a}_{\bs{q}}}\psi(\bs{q})\Bigr)\nonumber\\
&=&\Bigl(\frac{1}{2}m\bs{q}^2+w+\frac{1}{2}m\dot{\bs{q}}\cdot\bs{a}_{\bs{q}}+i\dot{\bs{q}}\cdot\nabla\Bigr)\psi(\bs{q}),\label{4.2.11}
\end{eqnarray}
where the identity $\dot{\bs{a}_{\bs{q}}}=\bs{q}$ is used to obtain the first term of the third equality. We recognize  the first two terms as kinetic energy 
and internal energy while the second two terms are contributions due to acceleration. Aside from the second, all terms are functions of $t$, including  the kinetic energy term 
$\frac{1}{2}m\bs{q}(t)^2$, in contrast to the Galilei case. The internal energy, on the other hand, is not a function of $t$ and, by construction, an invariant under the Galilean line group.

Similarly, we may define the momentum operator as the generator of the spatial translations subgroup. If we set $b=0$ and 
$\bs{a}(t)=\bs{a}^0$, a constant, in \eqref{4.23} and \eqref{4.23b}, we obtain (cf.~\eqref{4.6})
\begin{equation}
\left(\hat{U}(I,\bs{a}^0,0)\psi\right)(\bs{q})=e^{im\bs{q}\cdot\bs{a}^0}\psi(\bs{q})\label{4.2.12}.
\end{equation}
With the definition 
\begin{equation}
\hat{P}_i:=-i\left.\frac{d\hat{U}(I,{a}^0_i,0)}{da^0_i}\right|_{a^0_i=0}\label{4.2.13},
\end{equation}
from \eqref{4.2.12} we obtain
\begin{equation}
\left(\hat{\bs{P}}\psi\right)(\bs{q})=m\bs{q}\psi(\bs{q})\label{4.2.14}.
\end{equation}
Since $\bs{q}(t)=\dot{\bs{a}_{\bs{q}}}$ is the velocity, \eqref{4.2.14} tells us that we may continue to define momenta $\hat{\bs{P}}$ as the generators of 
pure spatial translations $\bs{a}_0$ even in non-inertial reference frames.  This interpretation of momentum also suggests that we define position operators
\begin{equation}
\hat{\bs{X}}:=i\nabla_{\bs{p}}=i\frac{1}{m}\nabla_{\bs{q}}\label{4.2.15}
\end{equation} 
which are canonically conjugated to $\hat{\bs{P}}$: 
\begin{equation}
\left[\hat{X}_i,\hat{P}_j\right]=i\delta_{ij}\hat{I}\label{4.2.16}.
\end{equation}
 In terms of \eqref{4.2.14} and \eqref{4.2.15}, we can write the Hamiltonian \eqref{4.2.11} as 
 \begin{equation}
\left(\hat{H}\psi\right)(\bs{q})=\left(\frac{\hat{\bs{P}}^2}{2m}+\hat{V}+m\dot{\bs{q}}\cdot\left(\hat{\bs{X}}+\frac{1}{2}{\bs{a}_{\bs{q}}}\right)\right)\psi(\bs{q})\label{4.2.17}.
\end{equation}
Thus, unlike in the Galilei case, here $[\hat{H},\hat{\bs{P}}]\not=0$. 

For constant velocity $\bs{q}_0$, the Hamiltonian \eqref{4.2.17} reduces the case familiar from Galilean quantum physics: 
\begin{equation}
\left(\hat{H}\psi\right)(\bs{q}_0)=\left(\frac{\hat{\bs{P}}^2}{2m}+\hat{V}\right)\psi(\bs{q}_0)\label{4.2.18}.
\end{equation}

Note that in quantum mechanics  the Hamiltonian is often written as $\hat{H}=i\frac{d}{dt}$ and for a free particle it is equal to 
$\frac{\hat{\bs{P}}^2}{2m}$. However, we emphasize that the Hamiltonian, being the generator of time translations, must be defined as the derivative of the 
time translation group with respect to the \emph{translation parameter} $b$, evaluated at the group identity. 
Derivatives with respect to the 
coordinates of the underlying spacetime manifold do not have have a fundamental meaning as generators of the symmetry group on that manifold, although 
such relationships may be obtained as \emph{derived} expressions for certain choices of representation, cf.~\eqref{3.3.7}-\eqref{3.3.9}. The clear difference between 
$\frac{d}{db}$ and $\frac{d}{dt}$ must be borne in mind when dealing with the Hamiltonian \eqref{4.2.11} in quantum mechanics in non-inertial reference frames. In particular, 
while the Hamiltonian is a $t$-dependent operator, it is independent of $b$, the conjugate parameter with respect to which the exponentiation $e^{-i\hat{H}b}$ is done. That is 
to say that here we do not face the integrability issues, such as the general non-existence of $\hat{U}(t)=e^{-i\int_0^t dt'\hat{H}(t')}$, associated with the so-called time dependent Hamiltonians.

\section{Concluding remarks}\label{sec5}
We have presented an analysis of the Galilean line group, the  group of acceleration transformations that ties together all inertial and non-inertial reference frames. 
While both rotationally and linearly accelerating reference frames can be accommodated in this setting, it is only in the absence of former, it appears, 
that there exist extensions of the Galilean line group that contain central extensions of the Galilei group. Even under this restriction, the appropriate 
extensions of the Galilean line group are non-central. The key technical result we have reported in this paper is the unitary representations of the non-centrally 
extended Galilean line group, constructed under the restriction that there are no time dependent rotations. While these 
representations reduce to the usual projective representations of the Galilei group when restricted to transformations between inertial reference frames, 
they have a cocycle structure rather more complicated than that of projective representations. We postpone a more thorough study of the cohomology structure of the Galilean 
line group to a future paper. 

The representations of the Galilean line group presented here provides a means of understanding if the equivalence principle is consistent with quantum physics. 
In particular, from \eqref{4.2.17} we note that the Hamiltonian operator acquires the contribution 
$m\dot{\bs{q}}\cdot\left(\hat{\bs{X}}+\frac{1}{2}\bs{a}_{\bs{q}}\right)$ when transformed to a non-inertial reference frame. 
That this contribution is proportional to inertial mass is exactly what we would expect classical physics: fictitious forces an object experiences 
in non-inertial reference frames are 
proportional to its inertial mass. 

In the classical case, the proportionality of the fictitious forces to inertial mass immediately implies that the acceleration an object acquires due to the non-inertial 
character of the observer's reference frame is independent of its mass. Therewith, all objects will undergo the same changes to their trajectories in 
a non-inertial reference frame compared with the corresponding trajectories in an inertial reference frame.  In this regard, the quantum 
case is quite different from the classical case: the mass parameter in the fictitious potential term of \eqref{4.2.17} 
does not drop out in the Schr\"odinger's equation of motion and if we insert $\hbar$ explicitly in the Galilean line group representations 
(for convenience, we have set $\hbar=1$ in our construction), we see that the evolution of quantum state governed by \eqref{4.2.17} depends 
on the ratio $\frac{\hbar}{m}$. In other words, the effects due 
to the non-inertial nature of a reference frame will be measurably different for systems with different masses. However, this issue is quite separate from that of the equivalence 
of gravitational and inertial masses. 

The profound connection between gravitational fields and non-inertial reference frames in classical physics rests on the observation that accelerations and trajectories 
of all objects in a gravitational field are also independent of the inertial mass, leading to the inference that the ratio of gravitational mass (or, perhaps more precisely,  gravitational charge)  
$m_g$ to inertial mass $m$ is a universal constant.  Quantum mechanically, we expect the Hamiltonian of a non-relativistic quantum system in an external gravitational field 
to be of the form
\begin{equation}
\hat{H}=\frac{\hat{\bs{P}}^2}{2m}+\hat{V}+m_g\hat{\phi}_g\label{5.1},
\end{equation}
where $\hat{\phi}_g$ is the gravitational potential, written as a function of a suitable position operator. The comparison of \eqref{5.1} with \eqref{4.2.17} shows that if the universality 
of the ratio $\frac{m_g}{m}$ holds quantum mechanically as well, then the term $\dot{\bs{q}}\cdot\hat{\bs{X}}+\frac{1}{2}\dot{\bs{q}}\cdot\bs{a}_{\bs{q}}$ of \eqref{4.2.17} has 
physical interpretation as a quantum gravitational potential. It is appealing that it emerges naturally in our construction as a fundamental quantum mechanical result, as opposed to, say, 
the outcome of some sort of quantization of a pre-existing classical field. However, the appearance of the position operator $\hat{\bs{X}}$ (as well as time dependent space 
translation term $\bs{a}_{\bs{q}}$) in \eqref{4.2.17} shows that our construction gives rise only to gravitational potentials that are linear in position, though they may 
have arbitrary time dependence. In Appendix \ref{sec3.5}, we have briefly sketched a  possible approach to accommodate gravitational potentials with non-trivial spatial gradients, but much more work 
is needed to determine if this approach could lead to meaningful results. 

Conversely, a comparison between the dynamics governed by \eqref{4.2.17} and \eqref{5.1} may be used as a test of the universality 
of the ratio of gravitational mass to inertial mass. Note that in much the same way that the evolution of a quantum system in a non-inertial reference frame is not independent of inertial mass but 
depends on $\frac{\hbar}{m}$, 
the structure of \eqref{5.1} implies that the evolution of a quantum state in a gravitational field depends on $\frac{\hbar}{m}$ and $\frac{\hbar}{m_g}$, and not solely on the ratio $\frac{m_g}{m}$ necessarily. 
In fact, it is shown in \cite{kajari} that the wave vector  for such a state depends on $(mm_g)^{1/3}$ while in a certain semi-classical sense, the center of mass motion 
of certain wave packets indeed depends only on the ratio $\frac{m_g}{m}$.  Models known as atom trampoline \cite{kajari, kasevich,aminoff,ovchinnikov,wallis} 
and atom fountain \cite{kajari, kasevich2} aim to probe the ways in which the gravitational and inertial masses enter in quantum physics. While the evolution of a quantum system 
in a linear gravitational potential may depend on $m$ and $m_g$,  if this evolution is identical to the evolution in a suitable accelerating reference frame, 
then the ratio $\frac{m_g}{m}$ must be universal, i.e., the principle of equivalence between inertial and gravitational masses holds. The similarity of form of \eqref{5.1} and \eqref{4.2.17} suggests that this is so. 

\section*{Acknowledgments}
We acknowledge support from Grinnell College and Research Corporation. 
SW gratefully acknowledges insight from William Klink in  illuminating conversations.

%\appendix
%\appendixpage
%\addappheadtotoc
\section*{Appendix}\label{sec3.5}
{\bf The map group of the Galilei group from $\mathbb{R}\otimes\mathbb{R}^3$}\\
As seen from the discussion above, the representations of the Galilean line group allow us to describe quantum gravitational fields that, 
while they may be time dependent, are spatially constant. This is a direct consequence of the line group having been defined as the group of analytic 
functions of time to the Galilei group. As is the case in classical physics, accelerations that are functions of time alone are equivalent to 
gravitational fields that are functions of time alone. 

In order to accommodate spatially varying gravitational fields, we may consider the group of analytic functions from the entire spacetime 
manifold $\mathbb{R}^3\otimes\mathbb{R}$ to the Galilei group.  To that end, consider a set of functions $g=(R,\bs{a},b)$ defined by their action 
on the spacetime points as follows: 
\begin{equation}
(R,\bs{a},b):\ \ 
\left(\begin{array}{c}
\bs{x}\\
t
\end{array}\right)\rightarrow
\left(\begin{array}{c}
\bs{x}'\\
t'
\end{array}
\right)=
\left(\begin{array}{c}
R[\bs{x},t]\bs{x}+\bs{a}[\bs{x},t]\\
t+b[\bs{x},t]
\end{array}
\right)\label{3.5.1}
\end{equation}
In the interest of clarity, we use square brackets to denote evaluation of a function at a point.
The idea here is again rather similar to gauge theories and \eqref{3.5.1} can be viewed as gauging of the Galilei group itself. 

For ease of notation, we denote all four spacetime coordinates $(\bs{x},t)$ of a point  by $x$ and express \eqref{3.5.1} simply as
\begin{equation}
g: x \rightarrow x' =  g[x]x.\label{3.5.2}
\end{equation}
As in \eqref{3.1.4}, we define the composition of two elements $g'$ and $g$ by considering their combined action on a spacetime point $x$ given by \eqref{3.5.2}:
\begin{equation}
(g_2\circ g_1) x := g_2[x']x' = g_2 \big[g_1[x]x\big] g_1[x] x.  \label{3.5.3}
\end{equation}
On the right hand side, the multiplication $ g_2 \big[g_1[x]x\big] g_1[x]$ is the composition rule of the Galilei group, well-defined since both $g_1[x]$ and 
$g_2\big[g_1[x]x\big]$ are elements of the Galilei group for all spacetime points $x$. This operation can be performed after 
evaluating the action of function $g_1$ at the spacetime point $x$, $g_1[x]x$, given by \eqref{3.5.1}. 

In order to extract a composition rule from \eqref{3.5.3} under which the set of functions $g$ is closed,  
 we must be able to write $g_2\circ g_1$ as a single, well-defined function of $x$ with the range in the Galilei group. 
Much as the case was with the shift $\Lambda_b$ introduced for the purposes of making the closure manifest in the Galilean line group, 
here, too, we must define an operator $\Gamma$ on the set of functions \eqref{3.5.1} with analogous properties. To that end,  

let 
\begin{equation}
\Gamma_{g_1}\big(g_2[x]\big): = g_2 \big[g_1^{-1}[x]x\big]\label{3.5.4}
\end{equation}

We can then write \eqref{3.5.3} as 
\begin{equation}
 g_2\circ g_1 = (\Gamma_{g_1^{-1}}g_2)\cdot g_1,\label{3.5.5}
\end{equation}
where the dot on the right hand side denotes pointwise multiplication. That is, $(g_2\cdot g_1) [x] = g_2[x]g_1[x]$. 

Written explicitly in terms of $R$, $\bs{a}$, and $b$, the composition law \eqref{3.5.3} is
\begin{equation}
(R_2, \bs{a}_2, b_2)(R_1, \bs{a}_1, b_1) = \left(  (\Gamma_{g_1\inv}R_2)R_1, \  (\Gamma_{g_1\inv}R_2)\bs{a}_1 + (\Gamma_{g\inv}\bs{a}_2) , \  \Gamma_{g_1\inv}b_2+b_1  \right).\label{3.5.6}
\end{equation}

It follows from the definition \eqref{3.5.4} that the composition rule \eqref{3.5.5} is associative. More generally, if we were to not rely on the explicit form of $\Gamma_g$ given by 
\eqref{3.5.4}, then associativity would put constraints on the operators $\Gamma_g$. These can be obtained by 
using \eqref{3.5.5} to write the composition of these elements: 
\begin{eqnarray}
g_3\circ (g_2\circ g_1):\ x\rightarrow g_3[x']x' 
& =&g_3[(\Gamma_{g_1^{-1}}g_2)[x]\cdot g_1[x]x]\cdot(\Gamma_{g_1^{-1}}g_2)[x]\cdot g_1[x]x\nonumber\\
&=&(\Gamma_{(g_2\circ g_1)^{-1}}g_3)[x]\cdot(\Gamma_{g_1^{-1}}g_2)[x]\cdot g_1[x] x.\nonumber\\
\label{3.5.7}
\end{eqnarray} 
Then, associativity requires that 
\begin{eqnarray}
g_3\circ g_2\circ g_1:\  x \rightarrow\ x'
&=&(\Gamma_{g_1^{-1}}((\Gamma_{g_2^{-1}}g_3)\cdot g_2))[x]\cdot g_1[x]x \nonumber\\
&=&(\Gamma_{(g_2\circ g_1)^{-1}}g_3)[x]\cdot(\Gamma_{g_1^{-1}}g_2)[x]\cdot g_1[x]x,\nonumber\\
\label{3.5.8}
\end{eqnarray}
which gives us the condition
\begin{equation}
\Gamma_{g_1^{-1}}\Gamma_{g_2^{-1}}=\Gamma_{(g_2\circ g_1)^{-1}}.\label{3.5.9}
\end{equation}
This condition is obviously fulfilled by the operator defined by \eqref{3.5.4}. Therefore, under the composition rule \eqref{3.5.5}, the set of transformation elements \eqref{3.5.1} 
is a semigroup. 

If this set  is to be a group, we must also require that every $g=(R,\bs{a},b)$ have an inverse under the composition rule \eqref{3.5.5}. 
This means for every $g$ of the form \eqref{3.5.2},  there must  exist another $g^{-1}$ such that the following identity holds for every spacetime point $x$:
\begin{eqnarray}
(g\circ g\inv)[x]x&=&g[g\inv[x]x]\cdot g\inv[x]x\nonumber\\
& =&\left(\Gamma_{g_1}g_2\right)[x]x\nonumber\\
&=&x\label{3.5.10}.\nonumber\\
\end{eqnarray} 
That is, given any $g$, we need $g\inv$ such that every $x$ is a fixed point for $\Gamma_{g\inv} \cdot g\inv$. 
While we do not have a proof establishing the existence or the non-existence of such a $g^{-1}$ for every Galilei map group element $g=(R,\bs{a},b)$, 
considering how stringent the fixed point condition \eqref{3.5.10} is, we suspect that such a function $g^{-1}$ does not exist as an element of the set of $g$'s. 
Should this indeed be the case, then the set of transformations \eqref{3.5.1} is an infinite dimensional semigroup under the composition rule \eqref{3.5.6} 
and is not a group. 

\end{document}